\documentclass{JHEP3}

\usepackage{epsfig}
\usepackage{amsbsy}
\usepackage{varioref}
\usepackage{pifont}

\newcommand{\beq}{\begin{equation}}
\newcommand{\eeq}{\end{equation}}
\newcommand{\bea}{\begin{eqnarray}}
\newcommand{\eea}{\end{eqnarray}}
\newcommand{\nn}{\nonumber}
\def\beqn{\begin{eqnarray}} 
\def\eeqn{\end{eqnarray}} 

\def\eqn#1{Eq.~(\ref{#1})}
\def\eqns#1#2{Eqs.~(\ref{#1}) and~(\ref{#2})}

\def\fig#1{Fig.~{\ref{#1}}}
\def\vfig#1{Fig.~{\ref{#1}}}
\def\sec#1{Section~{\ref{#1}}}
\def\app#1{Appendix~\ref{#1}}

\def\spa#1.#2{\left\langle#1\,#2\right\rangle}
\def\spb#1.#2{\left[#1\,#2\right]}
\def    \sapp#1#2#3#4{{\langle #1 | (#2+#3) |#4  \rangle} }

\def\feynsl#1{
  \setbox0=\hbox{/} \setbox1=\hbox{$#1$}
  \dimen0=\wd0 \advance\dimen0 by -\wd1 \divide\dimen0 by 2
  \ifdim\wd0>\wd1 \raise.15ex\copy0\kern-\wd0\kern\dimen0\copy1\kern\dimen0
  \else \kern-\dimen0\raise.15ex\copy0\kern-\dimen0\kern-\wd1\copy1\fi}

\def\tr{\mathop{\rm tr}\nolimits}

\def\eff{{\rm eff}}
\def\gev{{\rm GeV}}
\def\ord{{\cal O} }
\def\cM{{\cal M}}
\def\cL{{\cal L}}
\def\la{\langle}
\def\ra{\rangle}

\def\ib{{\bar\imath}}
\def\jb{{\bar\jmath}}
\def\qb{{\bar q}}
\def\Qb{{\overline Q}}
\def\intdq{\int\kern-5pt{d^n\!q\over(2\pi)^n}}
\def\intdqfour{\int\kern-5pt{d^4\!q\over(2\pi)^4}}

\def\lq{\left[} 
\def\rq{\right]}

\def\({\left(} 
\def\){\right)} 
\def\de{\delta}

\newcommand\sss{\scriptscriptstyle}
\newcommand\as{\alpha_{\sss S}} 

\newcommand\h{{\rm H}}
\newcommand\ph{p_{\sss {\rm H}}}
\newcommand\php{p_{\sss {{\rm H}_\perp}}}

\newcommand\pbp{p_{b'\sss \perp}}
\newcommand\pip{p_{i\sss \perp}}
\newcommand\pjp{p_{j\sss \perp}}

\newcommand\qp{q_{\sss \perp}}
\newcommand\qip{q_{i\sss \perp}}
\newcommand\qap{q_{1\sss \perp}}
\newcommand\qbp{q_{2\sss \perp}}
\newcommand\mh{M_{\sss {\rm H}}}
\newcommand\mt{M_{t}}
\newcommand\pt{p_{{}\sss \perp}}
\newcommand\mhp{m_{\sss {{\rm H}_\perp}}}
\newcommand\mip{m_{i\sss \perp}}
\newcommand\yh{y_{\sss {\rm H}}}
\newcommand\sh{s_{j\sss {\rm H}}}
\newcommand\sah{s_{j_1\sss {\rm H}}}
\newcommand\sbh{s_{j_2\sss {\rm H}}}

\newskip\humongous \humongous=0pt plus 1000pt minus 100pt
\def\caja{\mathsurround=0pt}
\def\eqalign#1{\,\vcenter{\openup1\jot \caja
       \ialign{\strut \hfil$\displaystyle{##}$&$
        \displaystyle{{}##}$\hfil\crcr#1\crcr}}\,}
\newif\ifdtup

\newcommand{\ccaption}[2]{ 
  \begin{center} 
    \parbox{\textwidth}{ 
      \caption[#1]{\small
 {#2}}} 
  \end{center}    }

\newcommand\fverb{\setbox\pippobox=\hbox\bgroup\verb}
\newcommand\fverbdo{\egroup\medskip\noindent%
                        \fbox{\unhbox\pippobox}\ }
\newcommand\fverbit{\egroup\item[\fbox{\unhbox\pippobox}]}
\newbox\pippobox

\title{Kinematical Limits on Higgs Boson Production via Gluon Fusion 
in Association with Jets}

\author{V. Del Duca\\
Instituto Nazionale di Fisica Nucleare, Sez. di Torino\\
via P. Giuria, 1 - 10125 Torino, Italy\\
        E-mail: \email{delduca@to.infn.it}}

\author{W. Kilgore\\
Physics Department, Brookhaven National Laboratory\\
Upton, New York 11973, U.S.A.\\
	E-mail: \email{kilgore@bnl.gov}}

\author{C. Oleari\\
Institute for Particle Physics Phenomenology\\
University of Durham\\
South Road, Durham, DH1 3LE, U.K.\\
	E-mail: \email{carlo.oleari@durham.ac.uk}}

\author{C.R. Schmidt\\
Department of Physics and Astronomy,
Michigan State University\\
East Lansing, MI 48824, USA\\
	E-mail: \email{schmidt@pa.msu.edu}}

\author{D. Zeppenfeld\\
Department of Physics, University of Wisconsin\\
 Madison, WI 53706, U.S.A.\\
	E-mail: \email{dieter@pheno.physics.wisc.edu}}

\abstract{
In this paper, we analyze the high-energy limits for 
Higgs boson $+$ two jet production. We consider two high-energy limits,
corresponding to two different kinematic regions: (\emph{a})~the Higgs boson
is centrally located in rapidity between  
the two jets, and very far from either jet; (\emph{b})~the Higgs boson is
close to one jet in rapidity, and both of these  
are very far from the other jet. In both cases the amplitudes factorize
into impact factors or coefficient functions connected by gluons
exchanged in the $t$ channel. Accordingly, we compute
the coefficient function for the production of a Higgs boson from two 
off-shell gluons, and the impact factors for the production of a 
Higgs boson in association with a gluon or a quark jet.  We include the full
top quark mass dependence and compare this with the result obtained 
in the large top-mass limit.}
\keywords{Standard Model, QCD, jets, hadronic colliders, Higgs production}
\preprint{
~DCPT/02/148, ~DFTT 20/2002, ~IPPP/02/74\\ 
~MADPH~02-1276, ~MSUHEP-20620\\ 
}

\begin{document}

\section{Introduction}
\label{sec:intro}

In hadronic collisions,
Higgs boson production in association with two jets occurs through
gluon-gluon fusion and through weak-boson fusion (WBF).
The WBF process, $q\, q\to q\, q\, H$, occurs through the exchange of a
$W$ or a $Z$ boson in the $t$ channel, and is characterized by
two forward quark jets~\cite{Cahn,bjgap}.
This process will be crucial in trying to measure the
Higgs boson couplings~\cite{Zeppenfeld:2000td} at the 
Large Hadron Collider at CERN. In this respect, Higgs $+$ two jet production
via gluon-gluon fusion, which has a much larger production rate before cuts, 
can be considered a background: it
has the same final-state topology, and thus may hide the features of the
WBF process.  Luckily, the gluon-gluon fusion background can be reduced 
(but not eliminated) by 
requiring that the two jets are well-separated in rapidity and have a 
large invariant mass, 
$\sqrt{s_{j_1j_2}}$~\cite{DelDuca:2001eu,DelDuca:2001fn}.

The calculation of Higgs $+$ two jet production via gluon-gluon 
fusion is quite involved, even at leading order in $\alpha_s$,
because in this process the Higgs boson is
produced via a heavy quark loop. Triangle, box, and pentagon quark
loops occur, with by far the dominant contribution coming from the 
top quark.  However,
if the Higgs mass is smaller than the threshold for the creation of a 
top-quark pair, $\mh \lesssim 2 \mt$, the coupling of the Higgs to the
gluons via a top-quark loop can be replaced by an effective 
coupling~\cite{Shifman:1979eb,Ellis:1976ap}. That simplifies calculations
tremendously, because it effectively reduces the number of loops in a given 
diagram by one. We shall term calculations using the effective 
coupling as being done in the {\it large $\mt$ limit}.

In fully inclusive Higgs production, the condition $\mh \lesssim \mt$
is sufficient to use the large $\mt$ limit, because the only physical 
scales present in the production rate are the Higgs and the top-quark
masses. However, in Higgs production in association with one or more
jets, other kinematic invariants occur, like a Higgs-jet invariant mass or
a dijet invariant mass (if there are two or more jets), or other
kinematic quantities of interest, like the transverse energies of the
Higgs or of the jets. In this instance, the condition $\mh \lesssim \mt$
may not be sufficient to use the large $\mt$ limit.

Recently, we computed the scattering amplitudes and the cross
section for the production of a Higgs boson through gluon-gluon fusion,
in association with two jets, including the full $\mt$ 
dependence~\cite{DelDuca:2001eu,DelDuca:2001fn}, and have used it as a 
benchmark to test the large $\mt$ limit. We found that in order to
use the large $\mt$ limit, in addition to the necessary condition
$\mh \lesssim 2 \mt$, the jet transverse energies must be smaller
than the top-quark mass, $\pt \lesssim \mt$. That agrees with
the analysis of the large $\mt$ limit in the context of Higgs $+$ one jet
production~\cite{Baur}. However, we also found that
the large $\mt$ approximation is quite insensitive to the value of the
Higgs--jet and/or dijet invariant masses. This issue becomes important in
the context of its companion process: the isolation of Higgs production 
via WBF requires selecting on events with large dijet invariant mass. This cut
suppresses the gluon-gluon fusion contribution and reduces the QCD backgrounds.

In this paper, we consider in more detail the {\it high-energy limit}
for Higgs $+$ two jet production via gluon-gluon fusion.
To be precise, we term the high-energy limit to mean
the cases when Higgs--jet and/or dijet invariant masses become 
much larger than the typical momentum 
transfers in the scattering.  In addition to being crucial for the study of
Higgs-boson couplings through the WBF process, this limit is
also interesting {\it per se}. In fact, in the high-energy limit the
scattering amplitude factorizes into high-energy coefficient functions, or
{\it impact factors}, connected by a gluon exchanged in the $t$ channel.
We can obtain the amplitudes for different sub-processes by 
assembling together different impact factors.
Thus the high-energy factorization 
constitutes a stringent consistency check on any amplitude for
the production of a Higgs plus one or more jets.
In addition, the high-energy factorization is independent of the
large $\mt$ limit, {\it i.e.} the two limits must commute.

In the high-energy limit, one deals with a kinematic region 
characterized by two (or more) very different hard scales.
In the instances above, a large scale, of the order of the squared
parton center-of-mass energy  $s$, can be either
a Higgs--jet or a dijet invariant mass. A comparatively smaller scale,
of the order of a momentum transfer $t$, can be a jet (or Higgs)
transverse energy. If it is found that a fixed-order expansion 
of the parton cross section in $\as$ does not suffice to describe the 
data for the production rate under examination, then it may be necessary to
resum the large logarithms of type $\ln(s/|t|)$. This can be done
through the BFKL equation~\cite{Kuraev:1976ge,Kuraev:1977fs,
Balitsky:1978ic}, which is an equation for the (process independent)
Green's function of a gluon exchanged in the $t$ channel between
(process dependent) impact factors.

As a warm-up, we analyze in \sec{sec:hjet} the high-energy 
factorization in a simpler case:
Higgs $+$ one jet production in proton-(anti)proton, $pp$, scattering.
We obtain the impact factor for producing a lone Higgs and compare the 
high-energy limit with and without the large $\mt$ approximation.
In \sec{sec:h2jet}, we present the
amplitudes for $q\, Q\to q\, Q\, H$ and $q\, g\to q\, g\, H$ scattering,
already computed in Ref.~\cite{DelDuca:2001fn}, in a different form,
which is more convenient to analyze the high-energy limits.
In \sec{sec:helim}, we consider the high-energy factorization for 
Higgs $+$ two jet production, and analyze the two high-energy limits 
which can occur in this case: (\emph{a})~the Higgs boson centrally located in
rapidity between  
the two jets, and very far from either jet; (\emph{b})~with
the Higgs boson close to one jet in rapidity, and both of these 
very far from the other jet.  In the first case we compute
the coefficient function for the production of a Higgs from two 
off-shell gluons, and in the second case we compute the
impact factors for the production of a Higgs in association with a gluon 
or a quark jet.  Again we include the full
top quark mass dependence and compare this with the result obtained 
in the large $\mt$ limit.  Finally, in \sec{sec:conc}, we draw our conclusions.

\section{High-energy factorization of Higgs + one jet}
\label{sec:hjet}

The process $p p \to j H$ occurs through gluon-gluon fusion, with the Higgs
boson coupling to the gluons via a quark loop.  Representative amplitudes for
this process are given in \fig{fig:higgsjet}. All momenta are taken outgoing
and the relevant squared-energy scales are the parton center-of-mass energy
$s=(p_a+p_b)^2$, the momentum transfer $t=(p_a+p_H)^2$, the Higgs mass $\mh$,
the top-quark mass $\mt$ and the jet-Higgs invariant mass $\sh$. At leading order,
$s=\sh$ and $s + t + u = \mh^2$, where $u=(p_a+p_{b'})^2$.

The corresponding scattering amplitudes 
have been computed in Ref.~\cite{Ellis:1988xu}.

The high-energy limit for this process is given by $\sh\gg |t|, \mh^2$, where
the jet-Higgs mass is of the order of  the 
center-of-mass energy (equal at leading order), and is much larger than any other 
invariant. This implies that the rapidity interval between the jet and the 
Higgs is large.   This limit has been investigated before in Ref.~\cite{DelDuca:1994ga},
where the BFKL resummation of the logarithms of type $\ln(s/|t|)$ was performed.
We now review the results of this analysis in \sec{sec:helimjet}.

\subsection{The high-energy limit for Higgs + one jet}
\label{sec:helimjet}

We consider the production of a parton of momentum $p_{b'}$
and a Higgs boson of momentum $\ph$,
in the scattering between two partons of momenta $p_a$ and $p_b$.
The high-energy limit, $\sh \gg |t|,\, \mh^2$, is equivalent to requiring 
a strong ordering of the light-cone components of the final-state momenta, 
{\it i.e.} the multi-Regge kinematics~\cite{Kuraev:1976ge}
\beq
\ph^+ \gg p_{b'}^+ \, ,\qquad
\ph^- \ll p_{b'}^- \, ,\label{mrkhj}
\eeq
where we have introduced light-cone coordinates $p^{\pm}= p_0\pm p_z $, and
complex transverse coordinates $p_{\perp} = p_x + i p_y$ (see \app{sec:appa}).
Equation~(\ref{mrkhj}) implies that the rapidities are ordered as
\beq
\yh \gg y_{b'} + \left| \ln{\mhp\over |\pbp|} \right|
\, ,\label{ymrkhj}
\eeq
with
$\mhp = \sqrt{\mh^2 + |p_{\sss {{\rm H}_\perp}}|^2}$ the Higgs 
transverse mass.

\EPSFIGURE{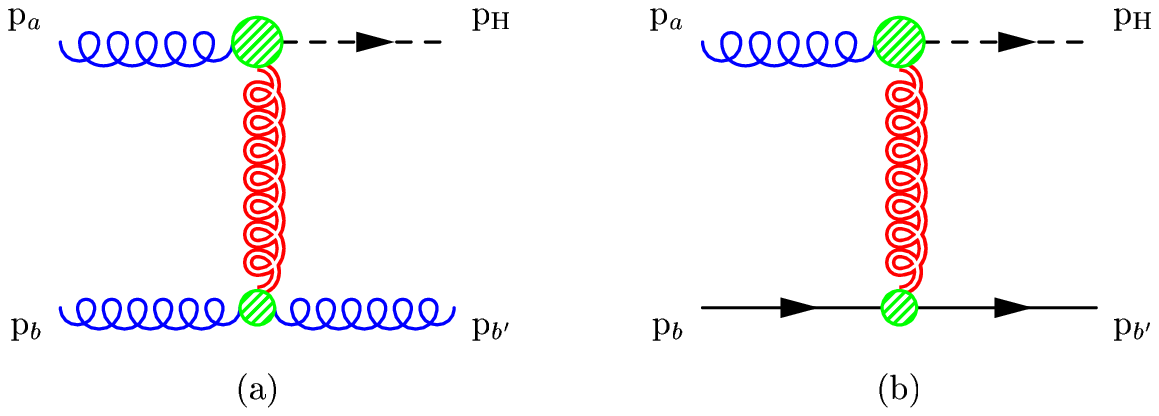,width=0.8\textwidth}
{$(a)$~Amplitudes for $g\, g \to H\, g$ scattering, and 
$(b)$~for $g\, q \to H\, q$ scattering
in the high-energy limit~(\ref{ymrkhj}).
The diagrams shown are meant to visualize the scattering amplitude only,
and have no direct relation to Feynman diagrams.
\label{fig:higgsjet} }

In the high-energy limit, the production rate is dominated by the parton
sub-processes which feature gluon exchange in the $t$ channel, represented
by a double curly line in \vfig{fig:higgsjet}.
The leading sub-processes are: $g\, g\to g\, H$, \fig{fig:higgsjet}~(a)
and $q\, g\to q\, H$, \fig{fig:higgsjet}~(b).
Since they only differ by the color strength in the jet-production
vertex, it is enough to consider one of them and include the others
through the effective parton distribution function
(p.d.f.)~\cite{Combridge:1984jn} 
\begin{equation}
f_{\rm eff}(x,\mu_F^2) = G\(x,\mu_F^2\) + {C_F\over C_A}\sum_f
\left[Q_f\(x,\mu_F^2\) + \bar Q_f\(x,\mu_F^2\)\right], \label{effec}
\end{equation}
where the sum runs over the quark flavours. In the high-energy limit,
the amplitude for $g\, g\to g\, H$ can be written in terms of an
effective vertex for the production of a gluon jet, $g^*\, g \rightarrow g$
(the lower blob in \fig{fig:higgsjet}~(a)),
and an effective vertex for the production of a Higgs boson, 
$g^*\, g \rightarrow H$ (the upper blob in \fig{fig:higgsjet}~(a)),
\beq
\label{HgggatHE}
i\ \cM^{gg\to Hg}\(p_a^{\nu_a}, \ph ;
p_{b'}^{\nu_{b'}}, p_b^{\nu_b}\) = 2\, s
\left[\delta^{ac} C^{g;\sss H}\(p_a^{\nu_a};\ph\)\right] {1\over t}
\left[i g\, f^{bb'c}\, C^{g;g}\(p_b^{\nu_b};p_{b'}^{\nu_{b'}}\) 
\right]\, ,
\eeq
with $q = p_b+p_{b'}$, $t\simeq - |\qp|^2$, and the $\nu$'s labelling 
the gluon helicities~\footnote{By convention, all particles
are taken as outgoing, thus an incoming quark (gluon) of a given helicity
is represented by an outgoing antiquark (gluon) of the opposite helicity.}. 
Using the high-energy factorization of the
amplitude, the effective vertices for $g^*\, g \rightarrow g$ can be
obtained from $g\, g\to g\, g$ scattering~\cite{DelDuca:1995zy}.
We define the impact factor as the square 
of each term in squared brackets in Eq.~(\ref{HgggatHE}),
summed (averaged) over final (initial) helicities and colors. Thus the
impact factor for the gluon jet is~\cite{Andersen:2001ja}
\begin{eqnarray}
\label{impgg}
I^{g;g}(p_b;p_{b'}) &=& \frac{1}{2(N_c^2-1)}
\left[i g\, f^{bb'c}\, C^{g;g}\(p_b^{\nu_b};p_{b'}^{\nu_{b'}}\) \right]
\left[- i g\, f^{bb'd}\, 
\left[C^{g;g}\(p_b^{\nu_b};p_{b'}^{\nu_{b'}}\)\right]^* \right]
\nonumber\\ &=& g^2\, {C_A\over N_c^2-1}\, \delta^{cd}\, ,
\end{eqnarray}
with $C_A=N_c$ = 3, and implicit sums over repeated indices. 
The impact factor for Higgs production is
\beq
\label{impgh}
I^{g;\sss H}(p_a;\ph) = \frac{1}{2(N_c^2-1)}
\left[\delta^{ac} C^{g;\sss H}\(p_a^{\nu_a};\ph\)\right]
\left[\delta^{ad} \left[C^{g;\sss H}\(p_a^{\nu_a};\ph\)\right]^* \right]\, .
\eeq
The squared amplitude, summed (averaged) over 
final (initial) helicities and colors, can be written as
\beq
\left|\cM^{g g\to H g}\right|^2 = {4s^2\over t^2}\,
I^{g;\sss H}(p_a;\ph)\, I^{g;g}(p_b;p_{b'})\, .\label{gg}
\eeq
Note that this squared amplitude also has a sum over the colors of the
gluons in the $t$-channel, the indices of which are implicitly included in the 
impact factors.
From the squared amplitude for $g\, g\to g\, H$ in the high-energy 
limit~\cite{DelDuca:1994ga}, we can extract the impact factor for
Higgs production
\beq
\label{imphiggs}
I^{g;\sss H}(p_a;\ph) = \frac{1}{N_c^2-1} \,{\as^2\over v^2}\, 
{1\over 128\pi^2} \, \delta^{cd} \,|\qp|^2\, 
\left|{\cal F}\(|\qp|^2\)\right|^2\, ,
\eeq
where $v$ is the vacuum expectation value parameter,
$v^2=(G_F\sqrt{2})^{-1}\simeq(246.22~\gev)^2$ and
${\cal F}(|\qp|^2)$ is given by~\cite{DelDuca:1994ga}
\bea
\label{formfac}
{\cal F}\(|\qp|^2\) &=& \({4\mt^2\over \mhp^2}\)
\left\{-2\,-\,\({2|\qp|^2\over \mhp^2}\)
\left[\sqrt{b}\,W(b)-\sqrt{a}\,W(a)\right] \right.\nonumber\\
&+& \left.{1\over 2} \(1-{4\mt^2\over \mhp^2}\)
\left[ W(b)^2-W(a)^2 \right]
\right\}\, ,
\eea 
with 
$$
W(c)\ =\ \cases{\phantom{\bigg[}\!
-2i\arcsin{(1/\sqrt{1-c})}\ ,&$\qquad\qquad
c<0$\cr\phantom{\bigg[}\!
\ln\displaystyle{1+\sqrt{c}\over1-\sqrt{c}}
\,-\,i\pi\ ,&$\qquad\qquad 0<c<1$\cr
\phantom{\bigg[}\!
\ln\displaystyle{\sqrt{c}+1\over\sqrt{c}-1}
\ ,&$\qquad\qquad c>1$\cr} \label{wform}
$$
and
\beq
a=1+{4\mt^2\over |\qp|^2}\, ,\qquad b = 1 - {4\mt^2\over \mh^2}\, ,
\eeq
and we take the root $\sqrt{b}=i\sqrt{|b|}$ for $b<0$.  

The amplitude for $q(\bar{q})\, g\to q(\bar{q})\, H$ scattering,
\fig{fig:higgsjet}~(b),
has the same analytic form as \eqn{HgggatHE}, up to
the replacement of an incoming gluon with a quark. If that occurs on the
lower line, we perform the substitution
\begin{equation}
\label{qlrag}
i g\, f^{bb'c}\, C^{g;g}\(p_b^{\nu_b};p_{b'}^{\nu_{b'}}\) \leftrightarrow g\, 
T^c_{b' \bar b}\, C^{\bar q;q}\(p_b^{-\nu_{b'}}; p_{b'}^{\nu_{b'}}\)\, ,
\end{equation}
and similar ones for an antiquark and/or for the upper line.  The effective
vertices for the production of a quark jet, $g^*\, q \rightarrow q$ are given
in Ref.~\cite{DelDuca:2000ha,DelDuca:1996km}. Then the impact factor for the
quark jet is~\cite{Andersen:2001ja}~\footnote{We use the standard
normalisation of the fundamental representation matrices, $\tr(T^a T^b) =
\delta^{ab}/2$ throughout.}
\begin{eqnarray}
I^{\bar q;q}(p_b;p_{b'}) &=& \frac{1}{2N_c} \left[g\
T^c_{b' \bar b}\, C^{\bar q;q}\(p_b^{-\nu_{b'}};p_{b'}^{\nu_{b'}} \)\right]
\left\{ g\ T^d_{\bar b b'}\, \left[
C^{\bar q;q} \(p_b^{-\nu_{b'}};p_{b'}^{\nu_{b'}}\)\right]^* \right\}
\nonumber\\ &=& g^2\, {1\over 2N_c}\, \delta^{cd}\, .\label{ifqq}
\end{eqnarray}
The ratio of the impact factor for the quark above to the one of
the gluon~(\ref{impgg}) is the color factor $C_F/C_A$
of the effective p.d.f.~(\ref{effec}).

\subsection{The combined high-energy limit and large top-mass limit}
\label{sec:hetoplimjet}

In the large top-mass limit, \eqn{formfac} reduces to~\cite{DelDuca:1994ga}
\beq
\lim_{|\qp|,\, \mh\, \ll\, \mt} {\cal F}\(|\qp|^2\) = - {4\over 3}
\eeq
and the impact factor $g^*\, g \rightarrow H$ for Higgs production becomes
\beq
\lim_{|\qp|,\,\mh\, \ll\, \mt} I^{g; \sss H}(p_a;\ph) = 
\frac{1}{N_c^2-1} {A^2\over 8}
\delta^{cd} \, |\qp|^2\, ,\label{imphigmtlim}
\eeq
where 
\beq
\label{eq:A_def}
A = \frac{\as}{3 \pi v}\, .
\eeq 

Alternatively, the impact factor can be obtained directly in the large $\mt$
limit, which is simplified by using a Lagrangian with an effective
Higgs-gluon-gluon operator, as discussed in \app{sec:toplim}.  In this
approach the constant $A$ above is just the coupling of the effective
operator (see Eq.~(\ref{efflag})).  The relevant sub-amplitudes in
the large $\mt$ 
limit are given in 
\app{sec:toplimhj}, and for Higgs plus three gluons, only the sub-amplitude
(\ref{mpp}) is leading in the high-energy limit.  Using the decomposition
(\ref{GluonDecompNew}), we can then write the amplitude for $g\, g\to H\, g$
in the form of \eqn{HgggatHE} with effective vertex $g^*\, g\to H$
\beq
C^{\sss H}(p_a^-;\ph) = -i {A\over 2\sqrt{2}} \,\qp\, .\label{imphigthe}
\eeq
Using \eqns{impgh}{imphigthe}, we obtain the same result~(\ref{imphigmtlim})
as above for the impact factor for Higgs production $g^*\, g\to H$ in the
large top-mass limit.  This verifies that 
the two limits commute.  It is easy to
check that the large $\mt$ amplitudes for $q\,g\rightarrow q\,H$ also
factorize in the high-energy limit as expected.


\subsection{The production rate for Higgs + one jet}
\label{sec:xsecthj}

%
%
%
%
%
%

Using the squared-matrix element formula~(\ref{gg}), it is straightforward
to compute the high-energy limit of the differential cross section for a
Higgs boson with an associated jet in hadron-hadron collisions.  For this one
must keep in mind that the Higgs boson rapidity may be either much larger
than the jet rapidity (as given by the limits~(\ref{mrkhj}) and
(\ref{ymrkhj})) or much smaller than the jet rapidity (obtained by reversing
the limits in~(\ref{mrkhj}) and~(\ref{ymrkhj})).

\begin{figure}[thb] 
\centerline{ 
\epsfig{figure=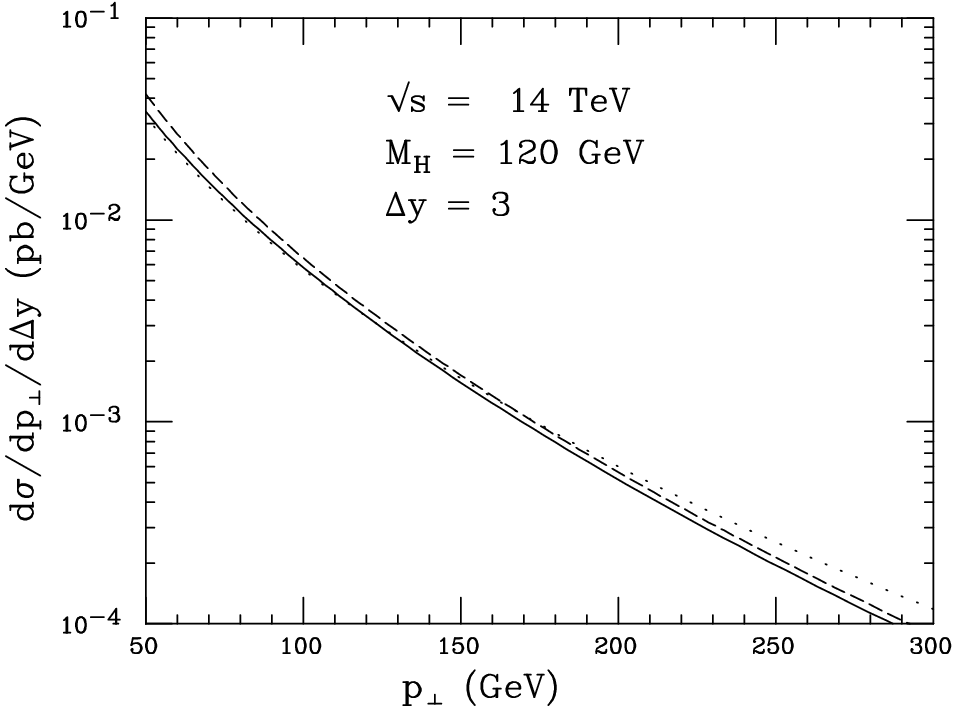,width=0.5\textwidth,clip=} \ 
\epsfig{figure=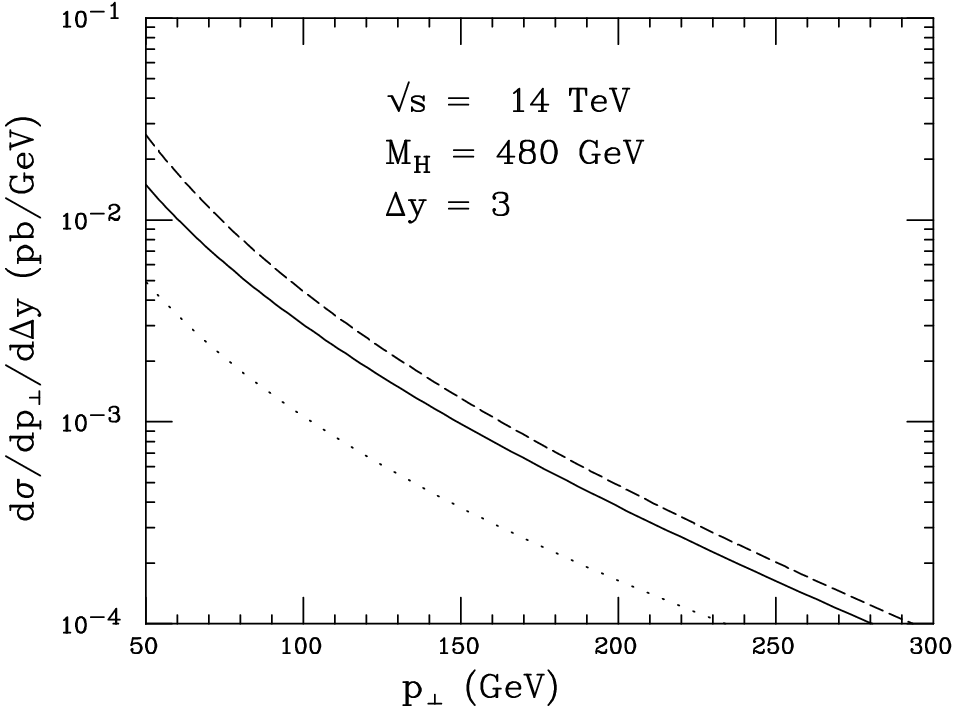,width=0.5\textwidth,clip=}}   
\ccaption{} 
{ \label{fig:ptdistra} 
Transverse momentum distribution of the Higgs boson (and of the jet) in $H$ +
one jet production in $pp$ collisions at the LHC energy $\sqrt{s}=14$~TeV.
The rapidity difference between the jet and the Higgs is taken to be $\Delta
y = 3$.  The Higgs mass has been fixed to
$\mh = 120$~GeV in the left panel, and to $\mh = 480$~GeV in the right panel.
The curves show the leading order amplitudes computed exactly (solid line)
with top mass $\mt = 175$~GeV,
in the $\mt\,\to\,\infty$ limit (dotted line) and in the high-energy limit
(dashed line). }
\end{figure}



In \vfig{fig:ptdistra}, we plot the transverse momentum distribution of the
Higgs boson (and of the jet) in $H$ + one jet production in $pp$ collisions at
the LHC energy $\sqrt{s}=14$~TeV. The rapidity difference between the jet and
the Higgs is taken to be $\Delta y = 3$. 
The top-quark mass has been fixed to $\mt = 175$~GeV while the
the Higgs mass has been chosen to be 
$\mh = 120$~GeV in the left panel, and $\mh = 480$~GeV in the right panel.
The curves in \fig{fig:ptdistra} (all at leading order in $\alpha_s$) 
correspond to the
exact evaluation of the production rate, according to the formulae of
Ref.~\cite{Ellis:1988xu} (solid line); to the rate with the amplitudes
evaluated in the $\mt\,\to\,\infty$ limit, as given in \sec{sec:toplimhj}
(dotted line); and to the rate evaluated in the high-energy limit, according
to the results of \sec{sec:helimjet}
(dashed line). 
We use the exact production rate as a benchmark
to which to compare the large $\mt$ limit and the high-energy limit. 

For the lower range of transverse momenta and for $\mh \lesssim 2\mt$ (left
panel), the large $\mt$ limit approximates the exact calculation very well,
but begins to deviate from it for $p_\perp\gtrsim\mh$~\cite{Baur}.
Note that the large $\mt$ limit fares well even though $\mt$ is not the
largest kinematic invariant. In fact, for $\mh = 120$~GeV, and $p_\perp >
50~$GeV, the jet-Higgs invariant mass is $\sqrt{\sh} \gtrsim 390$~GeV. This
behavior is confirmed from the analysis of the transverse momentum
distribution at larger values of $\Delta y$: the large $\mt$ limit is
insensitive to the value of the jet-Higgs invariant mass.  This is not
unexpected, since at large values of $\Delta y$ the cross section is
effectively dominated by diagrams with gluon exchange in the $t$ channel, as
displayed in \fig{fig:higgsjet}. Thus it is well described by the high-energy
factorization of \sec{sec:helimjet}. When, in addition, we take the large
$\mt$ limit, this does not modify the high-energy factorization, but only the
impact factor for Higgs production, as in \sec{sec:hetoplimjet}. In the
high-energy limit, the sensitivity to the full $\mt$-dependence does not
occur globally at the level of the entire amplitude, but locally (in
rapidity) at the level of the impact factor for Higgs production.

For $\mh \gtrsim 2\mt$ (right panel), the large $\mt$ limit does no longer
approximate the exact solution, while the high-energy limit only
gets slightly worse, due to the $\mh$ dependence in \eqn{ymrkhj}, over the
entire range of transverse momenta and for this rapidity separation.



\section{Exact amplitudes for Higgs + two jets}
\label{sec:h2jet}

We are interested in the high-energy limit of Higgs + two jet production in
$pp$ scattering, $pp\rightarrow j_1j_2H$, via gluon-gluon fusion. We computed
the exact leading order amplitudes, including the full dependence on the 
top quark mass $\mt$, in Ref.~\cite{DelDuca:2001fn}.

In this section we give analytic formulae for the amplitudes for $q\,Q\to
q\,Q\,H$ and $q\,g\to q\,g\,H$ scattering in an alternate form that is more
suitable for extracting the high-energy limit.  We shall see that these
amplitudes are sufficient to obtain the relevant production vertices and
impact factors in the high-energy limit.  Because of their complexity, we do
not give the expressions for the $g\,g\to g\,g\,H$ here, although in
principle, they could be used as an analytic cross-check.  Instead we have
performed this cross-check numerically.

\subsection[The amplitude for ${q\,Q\to q\,Q\,H}$ scattering]
{The amplitude for $\boldsymbol{q\,Q\to q\,Q\,H}$ scattering}

\label{sec:hqqqq}

There is one Feynman diagram for $q\,Q\to q\,Q\,H$ scattering, obtained by
inserting a $g\,g\,H$ triangle-loop coupling~\footnote{In counting diagrams,
we exploit Furry's theorem and count as one the two charge-conjugation
related diagrams where the loop momentum runs clockwise and
counter-clockwise.}  on the gluon exchanged in the $t$ channel of the
corresponding $q\,Q\to q\,Q$ diagram. The color decomposition of the
amplitude is identical to the corresponding QCD amplitude
\begin{equation}
\cM\(1_q, 2_{\bar{q}}; 3_Q, 4_{\Qb}\) = g^2 
\(T^a\)_{i_1}^{\ib_2}\, \(T^a\)_{i_3}^{\ib_4}\,
    m\(1_q, 2_{\bar{q}}; 3_Q, 4_{\Qb}\)\, .\label{FourQuarkDecomp}
\end{equation}

For simplicity we treat all momenta as outgoing (with 
$p_{H}+p_{1}+p_{2}+p_{3}+p_{4}=0$), and
we write the helicity amplitudes in terms of products of 
massless Weyl spinors $\psi_{\pm}(p)$ of fixed helicity
\begin{equation}
\psi_{\pm}(p) = {1\pm \gamma_5\over 2} \psi(p) \equiv 
|p^\pm\rangle\, , \qquad \overline{\psi_{\pm}(p)} \equiv \langle p^\pm|
\, .\label{spii}
\end{equation}
We use the following notation~\cite{Mangano:1990by} for
spinor products
\begin{equation}
\langle p k\rangle \equiv \langle p^- | k^+ \rangle\, , \qquad 
\left[pk\right] \equiv \langle p^+ | k^- \rangle\, , \qquad {\rm with}\;\;
\langle p k\rangle^* = {\rm sign}(p^0 k^0) \left[ k p\right]\, ,\label{spino}
\end{equation}
currents
\bea
\langle i| k | j\rangle &\equiv&
\langle i^-| \slash  \!\!\! k  |j^-\rangle = 
\langle i k \rangle \left[k j\right]\, ,\nonumber\\
\langle i| (k+l) | j\rangle &\equiv& 
\langle i^-| (\slash  \!\!\! k + \slash  \!\!\! l ) |j^-\rangle
= \langle i| k | j\rangle + \langle i| l | j\rangle\, 
,\label{currentsi}
\eea
and Mandelstam invariants
\begin{equation}
s_{pk} = 2\, p\cdot k = 
\langle p k \rangle \left[kp\right]\, . 
\end{equation}

There is only one independent sub-amplitude, which we obtain by saturating
the off-shell gluons from the triangle loop 
(\ref{Dj}) with two fermion currents
\bea
\lefteqn{i\, m\(1_q^+, 2_{\bar{q}}^-; 3_Q^+, 4_{\Qb}^-\) }\nonumber\\ &=& 
4\ {g^2\mt^2\over\ v}\ {\sapp2341\sapp4123 A_1(12; 34) - 
2\spb1.3 \spa4.2 A_2(12; 34)
\over s_{12} s_{34} }\, ,\label{hqqqq}
\eea
where the functions $A_{1,2}(12;34)\equiv A_{1,2}(p_{1}+p_{2};\, p_{3}+p_{4})$
are the two independent 
form factors from the triangle loop, defined in \eqn{a1a2}.
They are related to the form factors $F_T$ and $F_L$ of 
Ref.~\cite{DelDuca:2001fn} by
\beq
A_1(q_1,q_2) = {i\over (4\pi)^2} F_T\;, \qquad
A_2(q_1,q_2) = {i\over (4\pi)^2} \left( F_T\ q_1\cdot q_2 + F_L\ q_1^2 q_2^2
\right) \; .
\eeq

All other helicity configurations are related to \eqn{hqqqq} by 
parity inversion and charge conjugation, except for the $(+--+)$ 
configuration, which can be obtained
by inverting one of the two quark currents. 
The $Q\Qb$ quark current, for example, is inverted by re-labeling 3 and 4.
For identical quarks, we must subtract from Eq.~(\ref{FourQuarkDecomp})
the same term with the quarks (but not the anti-quarks) exchanged 
($1_{q}\leftrightarrow3_{Q}$).

\subsection[The amplitude for ${q\,g\to q\,g\,H}$ scattering]
{The amplitude for $\boldsymbol{q\,g\to q\,g\,H}$ scattering}
\label{sec:hqqgg}

The diagrams for $q\,g\to q\,g\,H$ scattering are obtained 
from the three corresponding $q\,g\to q\,g$ amplitudes by inserting
a $g\,g\,H$ triangle-loop coupling
on any gluon line or by replacing the three-gluon coupling
with a $g\,g\,g\,H$ box loop.  There are three distinct box diagrams
corresponding to the three distinct orderings of the gluon momenta.
This gives a total of 10 diagrams: 7 with triangles and 3 with boxes.
The color decomposition of the amplitude is identical to the 
corresponding QCD amplitude
\beq
\cM\(1, 2; 3_q, 4_{\qb}\) = g^2 \left[
\(T^{a_{1}}T^{a_{2}}\)_{i_3}^{\ib_4}
    m\(1, 2; 3_q, 4_{\qb}\)
+\(T^{a_{2}}T^{a_{1}}\)_{i_3}^{\ib_4}
    m\(2, 1; 3_q, 4_{\qb}\)\right]
    \, .\label{TwoQuarkTwoGlueDecomp}
\eeq

There are only three independent helicity sub-amplitudes, which we
take to be 
\break
$m(1^{+},2^{-};3_{q}^{+},4_{\qb}^{-})$,
$m(1^{-},2^{+};3_{q}^{+},4_{\qb}^{-})$ and
$m(1^{+},2^{+};3_{q}^{+},4_{\qb}^{-})$.  All others can be obtained by 
parity inversion, reflection symmetry, and charge conjugation.
As for the case of $q\,Q\to q\,QH$, we write the helicity amplitudes in 
terms of products of massless spinors.  This is obtained by representing
the polarization vector of an outgoing gluon of momentum $p$ 
as~\cite{Kleiss:1985yh,Xu:1987xb}
\beq
\epsilon^{\mu}_{\pm}(p,q)\ =\ \frac{\pm\langle 
p\pm|\gamma^{\mu}|q\pm\rangle}{\sqrt{2}\langle q\mp|p\pm\rangle}\, ,
\eeq
where the arbitrary reference momentum $q$ satisfies $q^{2}=0$ and $q\cdot
p\ne0$.  Gauge invariance guarantees that the amplitude is independent of
$q$, since the difference $\epsilon^{\mu}(p,q)-\epsilon^{\mu}(p,q^\prime)$ is
proportional to $p^{\mu}$.

As explained in Appendix~\ref{sec:boxloop}, the box loops can be parametrized
in terms of 14 form factors $H_i$. However, by choosing the polarization
vectors to be $\epsilon(p_1,p_2)$ and $\epsilon(p_2,p_1)$, we can reduce the
dependence to only 6 form factors, each of which is related to another by
exchange of gluon momenta $p_1$ and $p_2$.  Other choices of reference
momenta will give expressions for the amplitudes involving some or all of the
other form factors: by gauge invariance they will reduce to the same
answer when expanded in terms of scalar integrals.  We have verified this
analytically for other gauge choices.  In this paper we just give the
simplest expressions in terms of the 6 form factors.  They are
\bea
\label{apmii}
i\,m\(1^+, 2^-; 3_q^+, 4_{\qb}^-\) &=& - 
{g^4\mt^2\over\ v}\,\frac{1}{s_{12}s_{34}}\ \Bigg\{ \nonumber\\
&&\phantom{+} \frac{\langle 2|3|1\rangle}{\langle13\rangle[24]}
\Big[
2s_{24}s_{13}\(H_4+H_5\)-s_{13}\Delta H_{12}\nonumber\\
&&\qquad\qquad
+\(s_{23}
s_{13}-s_{24}s_{14}\)\(s_{24} H_{10} -s_{13} H_{12}\)   \nonumber\\
&& 
\qquad\qquad-4\(s_{24}\)^2 A_1(2; 134)-4s_{13}s_{14}A_1(1; 234)
\Big]\nonumber\\
&&+\frac{\langle 2|4|1\rangle}{\langle13\rangle[24]}\Big[
2s_{24}s_{13}\(H_4+H_5\)-s_{24}\Delta H_{10}\nonumber\\
&&\qquad\qquad
+\(s_{23}
s_{13}-s_{24}s_{14}\)\(s_{24}H_{10}-s_{13}H_{12}\)\nonumber\\
&&\qquad\qquad+4s_{23}s_{24} A_1(2; 134)+4\(s_{13}\)^2A_1(1; 234)
\Big]\Bigg\}
\\
%
\label{ampii}
i\,m\(2^-, 1^+; 3_q^+, 4_{\qb}^-\) &=& -
i\,m\(1^+, 2^-; 3_q^+, 4_{\qb}^-\) \nonumber
\\
&&+
{4g^4\mt^2\over\ v}\ \Biggl\{
\frac{\langle 2|3|1\rangle}{\langle13\rangle[24]}
\frac{s_{13}}{s_{23}}A_1(1; 234)
-\frac{\langle 2|4|1\rangle}{\langle13\rangle[24]}
\frac{s_{24}}{s_{14}}A_1(2; 134)
\Biggr\}\nonumber\\
\\
%
\label{appii}
i\,m\(1^+, 2^+; 3_q^+, 4_{\qb}^-\) &=& - 
{g^4\mt^2\over\ v}\,\frac{1}{s_{34}}\ \Biggl\{\nonumber\\
&&\phantom{+} \frac{\langle 4|1|3\rangle}{\langle12\rangle\langle21\rangle}\Biggl[
-2s_{12}H_1+2s_{24}H_4+2s_{23}H_5-\Delta H_{12}\nonumber\\
&&\qquad\qquad-\biggl[
\frac{4(s_{12}+s_{23})^2}{s_{2\h}}\biggl(1+\frac{s_{34}}{s_{13}}
\biggr)+\frac{4s_{23}s_{24}}{s_{2\h}}
\biggr]A_1(2;134)\nonumber\\
&&\qquad\qquad+\biggl[
\frac{4(s_{12}s_{34}+s_{12}s_{24}-s_{23}s_{14})}{s_{1\h}}\biggl
(1+\frac{s_{34}}{s_{24}}
\biggr)\nonumber\\
&&\qquad\qquad\qquad-\frac{4s_{23}(s_{12}+s_{13})}{s_{1\h}}
\biggr]A_1(1;234)\nonumber\\
&&\qquad\qquad+2(s_{2\h}-s_{1\h})A_1({12};{34})-4A_2({12};{34})
\Biggr]\nonumber\\
&&+ \frac{\langle 4|2|3\rangle}{\langle12\rangle\langle21\rangle}\Biggl[
-2s_{12}H_2+2s_{13}H_4+2s_{14}H_5-\Delta H_{10}\nonumber\\
&&\qquad\qquad+\biggl[\frac{4(s_{13}\mh^2+s_{34}s_{23}-s_{12}s_{14})}
{s_{2\h}}-4s_{13}\biggr]A_1(2;134)\nonumber\\
&&\qquad\qquad+\biggl[
\frac{4s_{13}s_{14}}{s_{1\h}}\biggl(1+\frac{s_{34}}{s_{24}}
\biggr)+\frac{4(s_{12}+s_{13})^2}{s_{1\h}}
\biggr]A_1(1;234)\nonumber\\
&&\qquad\qquad-2(s_{1\h}-s_{2\h})A_1({12};{34})+4A_2({12};{34})
\Biggr]\Biggr\}\ .
\eea
As in the previous subsection, we write the arguments of the triangle form
factors by the number of the parton momentum, combining numbers if the
momenta are added, {\it e.g.} $A_{1,2}(1;234)\equiv
A_{1,2}(p_{1};\, p_{2}+p_{3}+p_{4})$.  The triangle and box form factors are
given in Appendices~\ref{sec:triangleloop} and~\ref{sec:boxloop},
respectively.  We have defined 
\bea s_{i\h}&=& (p_i + \ph)^2\nonumber\\
\Delta&=& s_{12}s_{34}-(s_{13}+s_{14})(s_{23}+s_{24})\ .\label{deti} 
\eea 
In Feynman graphs with an on-shell gluon attached to the top-quark triangle
we have also used the identity 
\beq 
\label{onemasstri}
A_2(1; 234)\ =\ \frac{1}{2}(s_{12}+s_{13}+s_{14})A_1(1; 234)\ ,
\eeq
as well as the similar identity with $1\leftrightarrow2$, to simplify the
expressions for the amplitudes.

\section{High-energy factorization of Higgs + two jets}
\label{sec:helim}

The relevant (squared) energy scales in the process $pp\rightarrow j_1j_2H$
via gluon-gluon fusion are the parton center-of-mass energy $s$, 
the Higgs mass $\mh^2$, the dijet invariant mass $s_{j_1j_2}$, and 
the jet-Higgs invariant masses $\sah$ and $\sbh$. At leading order they
are related through momentum conservation,
\beq
s = s_{j_1j_2} + \sah + \sbh - \mh^2\, .\label{Hjjmtmcons}
\eeq
There are two possible high energy limits to consider: 
%
\begin{enumerate}
\item $s_{j_1j_2}\gg\sah,\sbh\gg\mh^2$, \emph{i.e.} the Higgs boson is 
centrally located in rapidity between the two jets, and very far from 
either jet;
\item $s_{j_1j_2},\sbh\gg\sah,\mh^2$, \emph{i.e.} the Higgs boson is 
close to jet $j_1$ in rapidity, and both of these are very far from jet $j_2$. 
\end{enumerate}
%
In both cases the amplitudes will factorize into effective vertices connected 
by a gluon exchanged in the $t$ channel. The high-energy factorization 
constitutes a stringent consistency check on the amplitudes for
the production of a Higgs plus two jets.
In addition, the high-energy factorization is independent of the
large top-mass limit; therefore the two limits must commute.
We investigate these features of both possible high-energy limits in the next
sections. 

\subsection[The high-energy limit ${s_{j_1j_2} \gg \sah,\, 
\sbh\gg \mh^2}$]{The high-energy limit $\boldsymbol{s_{j_1j_2} \gg \sah,\, 
\sbh\gg \mh^2}$}

\label{sec:helimtwojet}

We consider the production of two partons of momentum $p_1$ and $p_3$ and a
Higgs boson of momentum $\ph$, in the scattering between two partons of
momenta $p_2$ and $p_4$, where all momenta are taken as outgoing.  In the
limit $s_{j_1j_2} \gg \sah,\, \sbh\gg \mh^2$ the Higgs boson is produced
centrally in rapidity, and very far from either jet.  Note that this limit is
a particular case of the more general limit which will be presented in
\sec{sec:limit2}. However, given its simplicity, we find it convenient to
display it first.  This limit is equivalent to requiring
\beq
p_1^+ \gg \ph^+ \gg p_3^+ \, ,\qquad
p_1^- \ll \ph^- \ll p_3^- \, ,\label{mrk}
\eeq
which entails that the rapidities are ordered as
\beq
y_3 + \left| \ln{\mhp\over |p_{3_\perp}|} \right| \ll \yh \ll
y_1 - \left| \ln{\mhp\over |p_{1_\perp}|} \right| \, .\label{ymrk}
\eeq
In this limit, the amplitudes are dominated by gluon exchange in the $t$
channel, with emission of the Higgs boson from the $t$-channel gluon, as
shown in \fig{fig:higgs2jets}.  Thus, we can simplify the sub-amplitudes of
\sec{sec:h2jet} accordingly, by evaluating the spinor products with the
formulae of \app{sec:appa}.

\EPSFIGURE{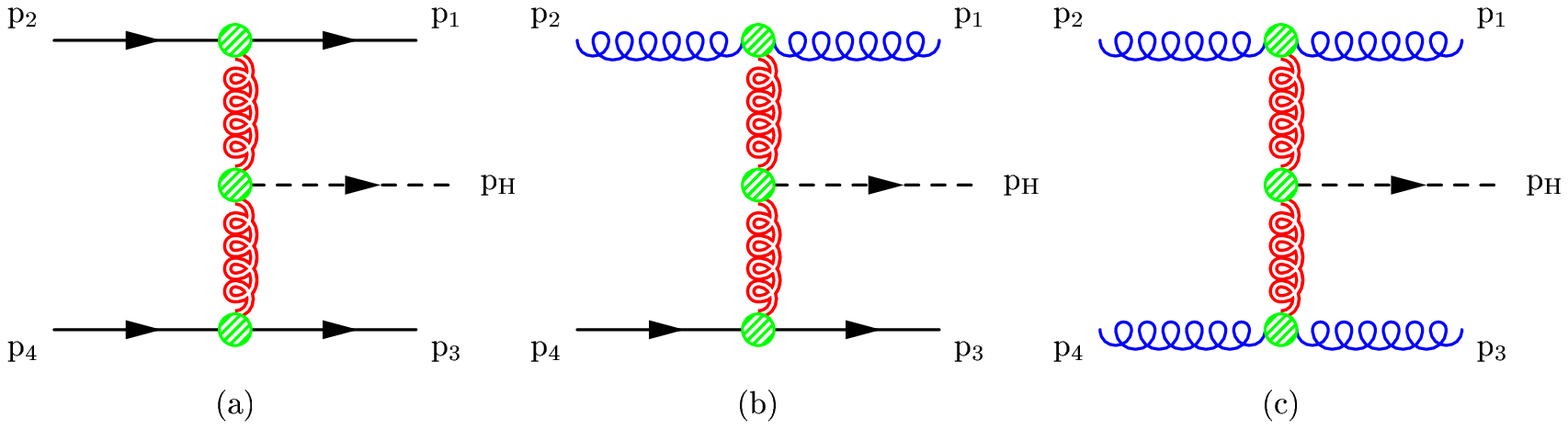,width=\textwidth}
{$(a)$ Amplitudes for $q\, Q \to q\, Q\, H$ scattering,
$(b)$ for $q\, g \to q\, g\, H$ scattering and $(c)$ for 
$g\, g \to g\, g\, H$ scattering, in the limit~(\ref{mrk}).
\label{fig:higgs2jets} }

For $q\, Q\to q\, Q\, H$ scattering, the amplitude, given by
\eqns{FourQuarkDecomp}{hqqqq}, and illustrated by \fig{fig:higgs2jets}~(a),
can be written in the high-energy limit as
\bea
\label{HqqqqHE}
\lefteqn{i\ \cM^{qq\to Hqq}\(p_2^{-\nu_1},\,p_1^{\nu_1} ,\, \ph,\,
p_3^{\nu_3},\, p_4^{-\nu_3}\) } \nonumber\\ 
&=& 2s \left[ g\,
T^c_{a_1 \bar a_2}\, C^{\bar q;q}\(p_2^{-\nu_1}; p_1^{\nu_1}\)\right]
{1\over t_1} \left[ \delta^{cc'} C^{\sss H}\(q_1,\ph,q_2\)\right] {1\over t_2}
\left[ g\, T^{c'}_{a_3 \bar a_4}\, 
C^{\bar q;q}\(p_4^{-\nu_3}; p_3^{\nu_3}\)\right]\!,
\eea
where $q_1 = - (p_1+p_2)$, $q_2 = p_3+p_4$, $t_i\simeq - |\qip|^2$,
$i = 1, 2$. In \eqn{HqqqqHE} we have made explicit the helicity conservation
along the massless quark lines. The effective vertex for Higgs production
along the gluon ladder, $g^*\, g^* \rightarrow H$,
with $g^*$ an off-shell gluon (represented by the central blob in
\fig{fig:higgs2jets}), is
\begin{equation}
\label{hif}
C^{\sss H}\(q_1,\ph,q_2\) = 2 \, {g^2 \mt^2\over v} \left[ \mhp^2 
A_1(q_1,q_2) - 2A_2(q_1,q_2) \right]\,,
\end{equation} 
with the coefficients $A_{1,2}$ defined in \eqn{a1a2}.  The effective
vertices $C^{\bar q;q}$ for $g^*\, q \rightarrow q$ and the ones for $g^*\,
\bar q \rightarrow \bar q$ were computed in
Refs.~\cite{DelDuca:2000ha,DelDuca:1996km}.

A similar analysis can be performed for $q\, g\to q\, g\, H$ scattering,
\fig{fig:higgs2jets}~(b), with only the sub-amplitudes~(\ref{apmii})
and~(\ref{ampii}) contributing in the high-energy limit.  This amplitude can
also be written as \eqn{HqqqqHE}, provided we substitute a gluon effective
vertex for one of the quark effective vertices as in~(\ref{qlrag}), where the
gluon effective vertex, $C^{g;g}$ was computed in Ref.~\cite{DelDuca:1995zy}.
We have verified
this analytically.  The same check on the (squared) amplitude for $g\, g\to
g\, g\, H$ scattering, \fig{fig:higgs2jets}~(c), has been performed
numerically. Thus, in the high-energy limit the amplitudes for $q\, Q\to q\,
Q\, H$, $q\, g\to q\, g\, H$ and $g\, g\to g\, g\, H$ scattering only differ
by the color strength in the jet-production vertex, and it is enough to
consider one of them and include the others through the effective
p.d.f.~(\ref{effec}).

Defining the coefficient function $V^{\sss H}$ for $g^*\, g^* \rightarrow H$,
\begin{equation}
\label{hifsq}
V^{\sss H}\(q_1,\ph,q_2\) = \left[ \delta^{cc'} C^{\sss H}(q_1,\ph,q_2) \right]
\left[ \delta^{dd'} \left[C^{\sss H}(q_1,\ph,q_2)\right]^* \right]\, ,
\end{equation} 
and using the impact factors for the gluon jets~(\ref{impgg}),
we can write the squared amplitude $g\, g\to g\, g\, H$, 
summed (averaged) over final (initial) helicities and colors, as
\bea
\label{Hgg}
\left|\cM^{gg\to Hgg}\right|^2 &=& 4 \,s^2\, I^{g;g}(p_2;p_1) \,
{1\over t_1^2}\, V^{\sss H}\(q_1,\ph,q_2\)\, {1\over t_2^2}\,
I^{g;g}(p_4;p_3)\nonumber\\ 
&=& 4 \, {C_A^2\over N_c^2-1}\, g^4\, { s^2\over t_1^2 \,t_2^2}\, \left|C^{\sss
H}\(q_1,\ph,q_2\)\right|^2\, .  
\eea

\subsubsection{The combined high-energy limit and large top-mass limit}
\label{sec:hetoplim}

These scattering amplitudes can be simplified further in the large top-mass
limit.  Taking this limit on the coefficients $A_{1,2}$ in the
vertex~(\ref{hif}), using~(\ref{a1a2t}) we obtain
\begin{equation}
\label{hifltm} 
\lim_{\mt\to \infty} C^{\sss H}\(q_1,\ph,q_2\) = 
i\ {A\over 2} \left( \left|\php\right|^2 - \left|\qap\right|^2 -
\left|\qbp\right|^2 \right)\, ,
\end{equation} 
with the effective coupling $A$ defined in Eq.~(\ref{eq:A_def}).
Note that there is a radiation zero when the transverse momenta 
are at right angles to each other~\cite{Plehn:2001nj}, giving the relation, 
$|\php|^2 = |\qap|^2 + |\qbp|^2$.
In the soft Higgs limit, $\ph\to 0$, the amplitudes are proportional to
those for dijet production in the high-energy limit.

We can obtain the same results by starting directly from the amplitudes for
Higgs plus two partons in the large $\mt$ limit, given in \app{sec:toplim}.
It is straightforward to show that in the high-energy limit they reduce to
the form of \eqn{HqqqqHE} (substituting gluon for quark effective vertices,
where appropriate, via~(\ref{qlrag})), with the effective vertex for the
Higgs given by \eqn{hifltm}.  Note that only the sub-amplitude~(\ref{hqqqqt})
for $q\, q\to q\, q\, H$ scattering, the sub-amplitudes~(\ref{qpmmp})
and~(\ref{qpmpm}) for $q\, g\to q\, g\, H$ scattering and the
sub-amplitude~(\ref{mmpp}) for $g\, g\to g\, g\, H$ scattering contribute 
in the
high-energy limit, because they are the only ones to allow conservation of
the quark and gluon helicities in the scattering.  Thus, we again find that
the two limits are interchangeable in all the sub-processes that survive 
in the high-energy limit~(\ref{mrk}).

\subsection[The high-energy limit ${s_{j_1j_2},\, \sbh\gg \sah,\,
\mh^2}$]{The high-energy limit $\boldsymbol{s_{j_1j_2},\, \sbh\gg \sah,\,
\mh^2}$}    
\label{sec:limit2}

Next, we consider the limit in which the Higgs boson is produced forward in
rapidity, and close to one of the jets, say to jet $j_1$, and both are very
far from jet $j_2$, {\it i.e.}  $s_{j_1j_2},\, \sbh\gg \sah,\, \mh^2$.
Labeling the partons as in the previous section, we find that this limit
implies
\beq
p_1^+ \simeq \ph^+ \gg p_3^+ \, ,\qquad
p_1^- \simeq \ph^- \ll p_3^- \, .\label{mrk2}
\eeq
In this limit, the amplitudes are again dominated by gluon exchange in 
the $t$ channel, and factorize into an effective vertex for the
production of a jet and another for the production of a Higgs plus a jet.

\EPSFIGURE{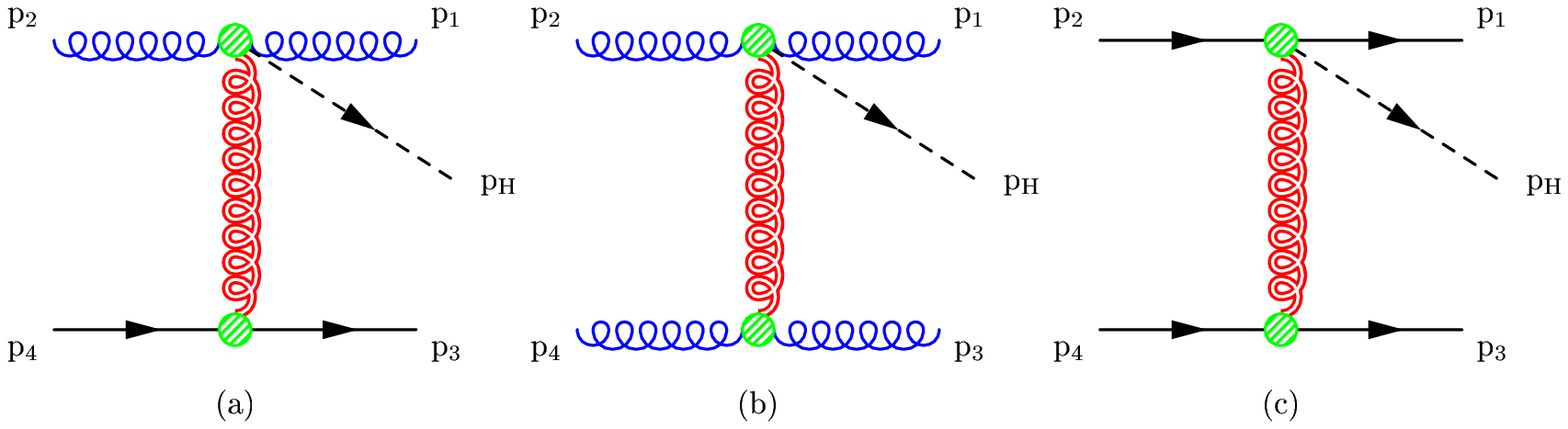,width=\textwidth}
{$(a)$~Amplitudes for $q\, g \to q\, H\, g$ scattering,
$(b)$~for $g\, g \to g\, H\, g$ scattering, and $(c)$~for 
$q\, q \to q\, H\, q$ scattering, in the limit~(\ref{mrk2}).
\label{fig:h2jetfigb} }

We consider first the amplitude for $q\, g\to q\, g\, H$ scattering
with the incoming gluon (quark) of momentum $p_2$ ($p_4$), as in
\vfig{fig:h2jetfigb}~(a). In this case
all of the sub-amplitudes~(\ref{apmii})--(\ref{appii}) are leading,
and to the required accuracy the sub-amplitude~(\ref{ampii}) is equal
and opposite to \eqn{apmii}. Then the amplitude for $q\, g\to q\, g\, H$ 
scattering can be written as
\bea
\label{HqgqgHE}
\lefteqn{ i\ \cM^{gq\to gHq}\(p_2^{\nu_2},\, p_1^{\nu_1}, \,\ph, \,
p_3^{\nu_3},\, p_4^{-\nu_3}\) } \phantom{aaaaaa} \nn\\ 
&=& 2 \, s \left[i g\, f^{a_2a_1c}\, 
C^{g; {\sss H} g}\(p_2^{\nu_2}; p_1^{\nu_1},\ph\)\right] 
{1\over t} \left[i g\, T^c_{a_3 \bar a_4}\, 
C^{\bar q;q}\(p_4^{-\nu_3};p_3^{\nu_3}\) \right]\, ,
\eea
where the effective vertex for the production of a quark jet 
$C^{\bar q;q}$ was computed in Refs.~\cite{DelDuca:2000ha,DelDuca:1996km},
and where 
$C^{g; {\sss H} g}(p_2^{\nu_2}; p_1^{\nu_1},\ph)$ is the
effective vertex for the production of a Higgs boson and a gluon jet,
$g^* g\to g H$ (the upper blob in \fig{fig:h2jetfigb}~(a) and~(b)). 
In \eqn{HqgqgHE} the momentum transfer is $t = q^2 \simeq
- |q_{\sss\perp}|^2$, with $q = p_3 + p_4$. 
There are two distinct helicity configurations for the effective vertex
$C^{g; {\sss H} g}$  that contribute.
For the helicity-conserving configuration we find
\bea
\lefteqn{ C^{g; {\sss H} g}(p_2^-; p_1^+,\ph) = - i
{g^4\mt^2\over\ v}\,\frac{1}{4 s_{12}}\ \frac{1}{s^2} 
\sqrt{1 + {\ph^+\over p_1^+}} } \phantom{aaaaaa}
\nonumber\\ && \Bigl\{ 
\Bigl( \langle 2|3|1\rangle + \langle 2|4|1\rangle\Bigr)\Bigl[
(s_{13})^2\Bigl(8A_1(1; 234)-2S_2H_{12}\Bigr)\nonumber\\
&&\qquad +(s_{24})^2\Bigl(-8A_1(2; 134)-2S_1H_{10}\Bigr)\nonumber\\
&&\qquad +s_{13}s_{24}\Bigl(4(H_4+H_5)+2S_2H_{10}+2S_1H_{12}\Bigr)
\Bigr]\nonumber\\
&& +
\Bigl( \langle 2|3|1\rangle - \langle 2|4|1\rangle\Bigr)\Bigl[
s_{13}\Bigl(-\Delta H_{12}-4S_1A_1(1; 234)\Bigr)\nonumber\\
&&\qquad+s_{24}\Bigl(\Delta H_{10}-4S_2A_1(2; 134)\Bigr)
\Bigr]\Bigr\}\, ,\label{coeffhggpm}
\eea
with invariants given in \eqn{invariants}, and currents in \eqn{currents}.
For the sake of compactness, we have relegated the necessary sub-formulae,
along with some details of the derivation, to \app{sec:vertex2}.  Note that
since the term in curly brackets is $\ord(s^2)$, \eqn{coeffhggpm} is
manifestly $\ord(s^0)$.

For the helicity-nonconserving configuration we find
\bea
C^{g; {\sss H} g}(p_2^+; p_1^+,\ph) &=& - i\ 
{g^4\mt^2\over\ v}\,\frac{1}{2 s_{12} }\ {1\over s}\
{p_{1_\perp}^*\over p_{1_\perp} } {p_{3_\perp} \over |p_{3_\perp}| }
\nonumber\\ && \biggl\{ - \langle 4|1|3\rangle\ 
(2s_{12}H_1 + \Delta H_{12}) 
- \langle 4|2|3\rangle\ (2s_{12}H_2 + \Delta H_{10}) \nonumber\\ && \quad
+2 \( \langle 4|1|3\rangle s_{24} + \langle 4|2|3\rangle s_{13}\)\ H_4
\nonumber\\ && 
\quad + 2 \bigl[ \langle 4|1|3\rangle S_2 + \langle 4|2|3\rangle S_1
- (\langle 4|1|3\rangle s_{24} + \langle 4|2|3\rangle s_{13} ) \bigr]\
H_5 \nonumber\\ && 
+ {4\over s_{13} s_{2\h} } \biggl[ 
\left[ s_{13} (2s_{12} + S_2) - s_{24} s_{34} \right]
( \langle 4|1|3\rangle s_{24} + \langle 4|2|3\rangle s_{13} )
\nonumber\\ && \qquad + \left[ 2 s_{24} s_{34} (s_{12} + S_2 )
- s_{13} (s_{12} + S_2 )^2 \right]\ \langle 4|1|3\rangle
\nonumber\\ && \qquad + s_{13} (s_{34} S_2 - s_{12} S_1)
\langle 4|2|3\rangle \biggr] A_1(2;134) \nonumber\\ && 
+ {4\over s_{24} s_{1\h} } \biggl[
\left[ s_{24} (2s_{12} + S_1) - s_{13} s_{34} \right]
( \langle 4|1|3\rangle s_{24} + \langle 4|2|3\rangle s_{13} )
\nonumber\\ && \qquad + \bigl[ s_{24} \left[ s_{12} s_{34} +
(s_{12} + S_1) (s_{34} - S_2) \right] + s_{13} s_{34} S_2 \bigr]\
\langle 4|1|3\rangle \nonumber\\ && \qquad +
\left( s_{13} s_{34} S_1 + s_{24} s_{12}^2 \right) \langle 4|2|3\rangle
\biggr] A_1(1;234) \nonumber\\ && + 2 (S_1 - S_2)\
( \langle 4|1|3\rangle + \langle 4|2|3\rangle )\ A_1(12;34)
\nonumber\\ && - 4 
( \langle 4|1|3\rangle - \langle 4|2|3\rangle )\ A_2(12;34) \biggr\}
,\label{coeffhggpp}
\eea
with invariants given in \eqn{invariants}, and currents in
\eqns{curr2}{curr3}. The term in curly brackets in \eqn{coeffhggpp} is
$\ord(s)$, and thus this effective vertex is $\ord(s^0)$. Finally, note that
\eqns{coeffhggpm}{coeffhggpp} transform under parity into their 
complex conjugates
\begin{equation}
[C^{g; {\sss H} g}(\{k^{\nu}\})]^* = C^{g; {\sss H} g}(\{k^{-\nu}\})\, . 
\end{equation} 

As in \eqn{impgg}, we define the impact factor for the production of a Higgs
and a gluon jet, summed (averaged) over final (initial) helicities and
colors, as
\begin{eqnarray}
\label{ifggh}
\lefteqn{ I^{g; {\sss H} g}\(p_2; p_1,\ph\) }\nn\\
&=& \frac{1}{2(N_c^2-1)}
\left[i g\, f^{a_2a_1c}\, C^{g; {\sss H} g}\(p_2^{\nu_2}; p_1^{\nu_1},\ph\)
\right] \left[- i g\, f^{a_2a_1 d}\, 
\left[C^{g; {\sss H} g}\(p_2^{\nu_2}; p_1^{\nu_1},\ph\)\right]^* \right]
\nonumber\\ &=& g^2\, {C_A\over N_c^2-1}\, \delta^{cd}\
\left[ \left|C^{g; {\sss H} g}\(p_2^-; p_1^+,\ph\)\right|^2 +
\left|C^{g; {\sss H} g}\(p_2^+; p_1^+,\ph\)\right|^2 \right]
\, ,
\end{eqnarray}
where in the first line a sum over repeated indices is implicit.
Finally, the squared amplitude, summed (averaged) over final (initial) 
helicities and colors is
\beq
\label{ggh}
\left|\cM^{g q\to g H q}\right|^2 = {4s^2\over t^2}\,
I^{g; {\sss H} g}\(p_2; p_1,\ph\)\, I^{\qb;q}\(p_4;p_3\)\, ,
\eeq
with the impact factor for the quark jet given by \eqn{ifqq}.
High-energy factorization implies that the amplitude for
$g\, g\to g\, g\, H$ scattering, \fig{fig:h2jetfigb}~(b), also
can be put in the form~(\ref{HqgqgHE}),
up to the replacement of the incoming quark with a gluon 
via the substitution~(\ref{qlrag}).

The amplitude for $q\, Q\to q\, Q\, H$ scattering, \fig{fig:h2jetfigb}~(c), 
factorizes into
an effective vertex for the production of a Higgs and a quark jet,
$g^* q\to q H$, and the effective vertex for the production of a quark
jet only. Using \eqns{FourQuarkDecomp}{hqqqq}, it can be written as
\bea
\label{HqqqqHE2}
\lefteqn{i\ \cM^{q Q\to q H Q}\(p_2^{-\nu_1},\, p_1^{\nu_1},\, \ph,\,
p_3^{\nu_3},\, p_4^{-\nu_3}\) } \phantom{aaaaa}
\nonumber\\ &=& 
2s \left[ g\, T^c_{a_1 \bar a_2}\, 
C^{\bar q; {\sss H} q}\(p_2^{-\nu_1};p_1^{\nu_1},\ph\)\right] {1\over t}
\left[ g\, T^{c}_{a_3 \bar a_4}\, 
C^{\bar q;q}\(p_4^{-\nu_3};p_3^{\nu_3}\)\right]\, ,
\eea
with the vertex for the production of a quark jet, $C^{\bar q;q}$, computed
in Refs.~\cite{DelDuca:2000ha,DelDuca:1996km}, and the vertex $g^* q\to q\,
H$ (the upper blob in \fig{fig:h2jetfigb}~(c)) given by
\bea
\label{hqqif}
C^{\bar q; {\sss H} q}\(p_2^-;p_1^+,\ph\) &=&
- 2 i\ {g^2 \mt^2\over v} {1\over s_{12}}\ 
\(1+{\ph^+\over p_1^+}\)^{-1/2}
\nn\\
& \times&
\left\{ \left[ \mhp^2 + {\ph^+\over p_1^+} 
\left( |p_{1_\perp}|^2 + p_{1_\perp}^* q_{{}_\perp} \right) \right]
A_1(q_1,q_2) - 2A_2(q_1,q_2) \right\} ,\phantom{aaa}
\eea
with coefficients $A_{1,2}$ defined in \eqn{a1a2}. Note that in the
more restrictive limit~(\ref{mrk}), where the Higgs is also 
separated by a large rapidity interval from jet $j_1$, this 
vertex factorizes further into an
effective vertex for the production of the Higgs along the ladder,
\eqn{hif}, a gluon propagator $1/s_{12}$, and an effective vertex
for the production of a quark jet, $C^{\bar q; q}(p_2;p_1)$. Finally, note
that under parity \eqn{hqqif} transforms 
into its complex conjugate and changes sign,
\begin{equation}
[C^{\qb; {\sss H} q}(\{k^{\nu}\})]^* = - C^{\qb; {\sss H} q}(\{k^{-\nu}\})\, . 
\end{equation} 

The impact factor for the production of a Higgs and a quark jet is given by
\bea
\label{ifqqh}
I^{\bar q; {\sss H} q}\(p_2;p_1,\ph\) &=&
{1\over 2N_c} \left[ g\, T^c_{a_1 \bar a_2}\, 
C^{\bar q; {\sss H} q}\(p_2^{\nu_2};p_1^{\nu_1},\ph\) \right]
\left[ g\, T^d_{\bar a_2  a_1}\, 
\left[ C^{\bar q; {\sss H} q}\(p_2^{\nu_2};p_1^{\nu_1},\ph\)\right]^* 
\right] \nn\\
&=& g^2\ {1\over 2N_c}\ \delta^{cd} 
\left| C^{\bar q; {\sss H} q}\(p_2^-;p_1^+,\ph\) \right|^2\ ,
\eea
with $\nu_2 = - \nu_1$, and an implicit sum over repeated indices. 
Thus the squared amplitude for
$q\,Q\to q\,H\,Q$ scattering, summed (averaged) over final 
(initial) helicities and colors, is
\beq
\label{qqh}
\left|\cM^{q Q\to q H Q}\right|^2 = {4s^2\over t^2}\,
I^{\qb; {\sss H} q}\(p_2; p_1,\ph\)\, I^{\qb;q}\(p_4;p_3\)\, ,
\eeq
with the impact factor for the quark jet given by \eqn{ifqq}.

\subsubsection{The combined high-energy limit and large top-mass limit}
\label{sec:hetoplim2}

In the large top-mass limit, the amplitude for $q\, Q\to q\,Q\,H$
scattering~(\ref{HqqqqHE2}) can be simplified further by using the
limits~(\ref{a1a2t}) for the coefficients $A_{1,2}$.  The effective vertex 
for the production of a Higgs and a quark jet~(\ref{hqqif}) reduces to
\begin{equation}
\lim_{\mt\to \infty} C^{\bar q; {\sss H} q}\(p_2^-;p_1^+,\ph\) 
= {A\over 2} {1\over s_{12}}\ 
\(1+{\ph^+\over p_1^+}\)^{-1/2}
\left[ p_{1_\perp} p_{3_\perp}^* +
\left(1 + {\ph^+\over p_1^+}\right) p_{1_\perp}^* p_{3_\perp}
\right]\ .\label{hevltm}
\end{equation} 
Again we note that in the limit~(\ref{mrk}), for which the Higgs is separated by 
a large rapidity interval from jet $j_1$, \eqn{hevltm} factorizes into the
effective vertex for the production of the Higgs along the ladder
in the large $\mt$ limit,
\eqn{hifltm}, a gluon propagator $1/s_{12}$, and an effective vertex
for the production of a quark jet, $C^{\bar q; q}(p_2;p_1)$.
Conversely, this vertex~(\ref{hevltm}) can be derived by taking the
limit~(\ref{mrk2}) on the amplitude for Higgs plus two $q\bar q$
pairs~(\ref{hqqqqt}) in the large $\mt$ limit.

Given the algebraic complexity of \eqns{coeffhggpm}{coeffhggpp}, it is easier
to derive the large $\mt$ limit of the effective vertex for the production of
a Higgs and a gluon jet by directly taking the limit~(\ref{mrk2}) on the
large $\mt$ amplitudes for $q\, g\to q\, g\,H$ scattering.  Thus, from
\eqns{qpmmp}{qpmpm} we obtain the helicity-conserving configuration
\beq
\lim_{\mt\to \infty} C^{g; {\sss H} g}(p_2^-;p_1^+,\ph) = 
i\, {A\over 2} {1\over s_{12}}\ 
\(1 + {\ph^+\over p_1^+} \)^{-1}
\left[ p_{1_\perp} p_{3_\perp}^* + \left(1 + {\ph^+\over p_1^+}\right)^2
p_{1_\perp}^* p_{3_\perp} \right]\ .\label{hggpmlmt}
\eeq
We have verified that we obtain the same result by taking the
limit~(\ref{mrk2}) on the amplitudes for $g\,g\to g\,g\,H$ scattering,
\eqn{mmpp}. From \eqn{qpmpp} we obtain the helicity-nonconserving
configuration
%
%
\bea
\label{hggpplmt}
&&\hspace{-0.5cm}\lim_{\mt\to \infty} C^{g; {\sss H} g}\(p_2^+;p_1^+,\ph\) 
= - i \,{A\over 2} {1\over s_{12}} 
\left\{ 
 p_{1_\perp}^* p_{3_\perp}^* \left( {\ph^+\over p_1^+} \right)^2
\(1 + {\ph^+\over p_1^+} \)^{-1} \right.\nn\\
&&\left. +
\left[ {(\php^*)^2\over |\php|^2 + p_1^+ \ph^-} -
\(1 + {\ph^+\over p_1^+}\)^{-1}
\frac{1}{\mh^2 + 2 p_1\cdot \ph }
 \left[ p_{3_\perp}^* + \left(1 + {\ph^+\over p_1^+} \right)
p_{1_\perp}^* \right]^2  \right] 
s_{12}
 \right\}\ .\nn\\
&& 
\eea
In the more restrictive limit~(\ref{mrk}), \eqn{hggpplmt} is subleading as
expected, while \eqn{hggpmlmt} factorizes further into the effective vertex
for the production of the Higgs along the ladder in the large $\mt$ limit,
\eqn{hifltm}, a gluon propagator $1/s_{12}$, and an effective vertex for the
production of a gluon jet, $C^{g; g}(p_2;p_1)$.

\subsection{The production rate for Higgs + 2 jets}
\label{sec:hexsect}

Using the formulae for the squared matrix elements, given in the previous
sections, one can compute the cross section for Higgs boson production in
association with two jets, in either of the two high-energy limits.  In the
numerical work in this section, we focus on the more restrictive
limit~(\ref{mrk}), where the Higgs boson is central in the rapidity interval
and widely separated from both of the two jets.


\EPSFIGURE{first_kin_limit.eps,width=0.8\textwidth}
{Cross section in $H+2$~jet production in $pp$ collisions 
at the LHC energy $\sqrt{s}=14$~TeV as a function of $\Delta y$,
with $\mh= 120$~GeV and $\mt = 175$~GeV. The dijet invariant mass
fulfils the constraint $\sqrt{s_{j_1j_2}} \ge 600$~GeV.
The rapidity interval $\Delta y$
is defined as $\Delta y = {\rm min}(|y_{j_1} - \yh|, |y_{j_2} - \yh|)$,
with the kinematical constraint $y_{j_1} > \yh > y_{j_2}$.
The solid line represents the exact production rate; the dashed line
the rate in the high-energy limit.
\label{fig:dydistra} }

\EPSFIGURE{first_kin_limit_mt_inf.eps,width=0.8\textwidth}
{ Same as Fig.~\ref{fig:dydistra}, with the solid line representing
the production rate in the large $\mt$ limit, and the dashed line the
rate in the combined large $\mt$ plus high-energy limit.
\label{fig:dydistrb} }

In Figs.~\ref{fig:dydistra} and~\ref{fig:dydistrb} we plot the cross section
for $H+2$~jet production in $pp$ collisions at the LHC energy
$\sqrt{s}=14$~TeV, as a function of $\Delta y$, where $\Delta y = {\rm
min}(|y_{j_1} - \yh|, |y_{j_2} - \yh|)$, with the kinematical constraint
$y_{j_1} > \yh > y_{j_2}$.  That is, $\Delta y$ is defined as the smallest
rapidity difference among the final-state particles.  
In addition we have imposed the following set of cuts
\beq 
\label{eq:cuts_min}
|p_{j_\perp}|>20\;{\rm GeV}, \quad |y_{j_{1,2}}|<5,\quad y_{j_1}y_{j_2}<0,
\quad \sqrt{s_{j_1j_2}} \ge 600~{\rm GeV}, \quad R_{jj}>0.6
\eeq
where $R_{jj}$ describes the separation of the two partons in the rapidity
$y$ versus azimuthal angle plane
\beq 
R_{jj} = \sqrt{\Delta y_{jj}^2 + \Delta\phi_{jj}^2}\;.
\eeq
In the plots we used values of the Higgs and top-quark masses of $\mh= 120$~GeV
and $\mt = 175$~GeV, respectively. A fixed value of $\as = 0.12$ has been 
chosen, and the factorization scale has been taken to be 
$\mu_{\sss {\rm F}} = \sqrt{p_{j_{1\perp}} p_{j_{2\perp}}}$. 
The leading order p.d.f.'s of the CTEQ4L
package~\cite{cteq4} have been used. In \fig{fig:dydistra} the solid line
represents the production rate with the exact amplitudes, as evaluated in
Ref.~\cite{DelDuca:2001fn}, and the dashed line represents the rate in the
high-energy limit ({\it i.e.,} using the squared matrix element
of~(\ref{Hgg}) with the effective p.d.f.~(\ref{effec})); in
\fig{fig:dydistrb} the solid line represents the production rate with the
amplitudes evaluated in the large $\mt$ limit (given in \app{sec:toplimhjj}),
and the dashed line the rate in the combined large $\mt$ plus high-energy
limit from \sec{sec:hetoplim}.

In Figs.~\ref{fig:dydistra} and~\ref{fig:dydistrb}, the agreement 
between the curves is very good for $\Delta y\gtrsim 3$. 
The agreement for $\Delta y\lesssim 3$, where the high-energy limit is
not expected to be a good approximation, depends on the
exact details of how the high-energy limit is taken.
In calculating the dashed curve of
\fig{fig:dydistra}, we took the coefficients $A_{1,2}$ of \eqn{hif}
to be a function of $q_1^2$ and $q_2^2$ evaluated in the high-energy
limit, {\it i.e.,}
$q_1^2 = -q_{1\perp}^2$ and $q_2^2 = -q_{2\perp}^2$; however,
in \eqn{HqqqqHE} we evaluated the gluon
propagators $t_1$ and $t_2$ at the exact values $q_1^2$ and $q_2^2$.
This choice makes the high-energy limit curve of \fig{fig:dydistra} (and
respectively
the combined limit curve of \fig{fig:dydistrb}) converge faster to the
exact
curve (to the large $\mt$ limit curve) than using $t_1=-q_{1\perp}^2$
and
$t_2=-q_{2\perp}^2$. Of course, either choice for the value of the gluon
propagators is equally valid in the high-energy limit.
We can only ascertain \emph{a posteriori} which gives a better numerical
agreement with the exact calculation at finite $\Delta y$.

\section{Conclusions}
\label{sec:conc}

In this paper we have analyzed the high-energy limit for Higgs boson plus one
and two jet production.  For Higgs boson plus two jet production we
considered two versions of the high-energy limit, corresponding to two
different kinematic regions: (\emph{a})~$s_{j_1j_2}\gg\sah,\sbh\gg\mh^2$,
\emph{i.e.} the Higgs boson is centrally located in rapidity between the two
jets, and very far from either jet;
(\emph{b})~$s_{j_1j_2},\sbh\gg\sah,\mh^2$, \emph{i.e.}  the Higgs boson is
close to jet $j_1$ in rapidity, and both of these are very far from jet
$j_2$. In all cases the amplitudes factorize into impact factors or
coefficient functions connected by gluons exchanged in the $t$ channel. We
obtained the relevant impact factors, keeping the
full $\mt$ dependence, for all of the partonic subprocesses which contribute
in the high-energy limits.  In particular, we computed the impact factor for
the production of a Higgs boson (\ref{imphiggs}) from the high-energy limit
of Higgs $+$ one jet production; the coefficient function for the production
of a Higgs from two off-shell gluons~(\ref{hifsq}) from case (\emph{a}) in
Higgs $+$ two jet production; and the impact factors for the production of a
Higgs in association with a gluon~(\ref{ifggh}) or a quark ~(\ref{ifqqh}) jet
from case (\emph{b}) in Higgs $+$ two jet production.  This required the full
one-loop $\mt$-dependent $q\,Q\to q\,Q\,H$ and $g\,q\to g\,q\,H$ amplitudes,
which we have presented here in a new form, which is convenient for our
purposes.

In addition, we have investigated the interplay of the high-energy limit with
that of the large $\mt$ limit.  We have computed the same impact factors and
coefficient functions as above, but also in the large $\mt$ limit, and we
have shown that the imposition of the two different limits commutes.  This is
important, because it implies that the factorization of amplitudes at high
energies ensures that the large $\mt$ computations are not made invalid by
the presence of an overall large energy scale, as long as the typical
transverse energy scale is less than or of the order of the top mass.  This
effect was first noted in Ref.~\cite{DelDuca:2001fn}, where we ascertained
that for Higgs $+$ two jet production the necessary conditions to use the
large $\mt$ limit are that the Higgs mass is smaller than the threshold for
the creation of a top-quark pair, $\mh \lesssim 2 \mt$ and the jet transverse
energies are smaller than the top-quark mass, $\pt \lesssim \mt$.  However,
the size of the dijet invariant mass was irrelevant to the applicability of
the large $\mt$ limit.

In general, the calculation of Higgs production via gluon-gluon fusion in
association 
with one or more jets is quite complicated, even at lowest order in $\as$,
because in this process the Higgs boson is produced via a heavy quark loop.
In Higgs $+$ one jet production, triangle and box loops occur;
in Higgs $+$ two jet production, pentagon loops also occur,
and in Higgs $+\ n$ jet production, up to $(n+3)$--side
polygon quark loops occur. In Ref.~\cite{DelDuca:2001fn} we
computed Higgs $+$ two jet production, but the complexity of
the calculation discourages one from carrying on this path. For instance,
the evaluation of the radiative corrections at $\ord(\as)$ to 
Higgs $+$ two jet production would imply the calculation of up to
hexagon quark loops and two-loop pentagon quark loops, which are at
present unfeasible.  Fortunately, there are two types of limits in which
the calculations become simpler:
\begin{itemize}
\item[-] the large $\mt$ limit, in which the heavy quark loop is replaced 
by an effective coupling, thus reducing the number of loops in a
given diagram by one;
\item[-] the high-energy limits, in which the number of sides in the
largest polygon quark loop is diminished at least by one.
\end{itemize}
We have seen that both of these limits can be relevant to phenomenology and
they can give useful and complementary information, both as cross checks of
an exact calculation with the full $\mt$ dependence, and as the starting
point for a more detailed calculation in which a well-based simplifying
assumption is crucial.

\section*{Acknowledgments}

C.R.S.~thanks the INFN, sez.~di Torino, for its kind hospitality during
portions of this work and acknowledges the U.S. National Science Foundation 
under grant PHY-0070443.  This research was supported in part by the University
of Wisconsin Research Committee with funds granted by the Wisconsin Alumni
Research Foundation and in part by the U.~S.~Department of Energy under
Contract No.~DE-FG02-95ER40896.
C.O. thanks the UK Particle Physics and Astronomy Research Council for
supporting his researches.

\appendix

\section{Scalar integrals}
\label{sec:scalars}

We define the scalar integrals by
\beqn
\label{scalars}
  B_{0}(k) &=& \intdq{1\over\left(q^2-\mt^2\right)
      \left[(q+k)^2-\mt^2\right]}
      -\frac{i}{16\pi^{2}} \(4\pi\)^\epsilon \mt^{-2\epsilon} 
      \frac{\Gamma(1+\epsilon)}{\epsilon}\\
  C_0(p;k) &=&
      \intdqfour{1\over\left(q^2-\mt^2\right)\left[(q+p)^2-\mt^2\right] 
      \left[(q+p+k)^2-\mt^2\right]}\\
  D_0(p;k;v) &=&
      \intdqfour \left\{ 
\frac{1}{\left(q^2-\mt^2\right) \left[(q+p)^2-\mt^2\right] 
      \left[(q+p+k)^2-\mt^2\right]}\right.\nn\\
&& \qquad \qquad \qquad \times
      \left.\frac{1}{\left[(q+p+k+v)^2-\mt^2\right]} \right\}\ . 
\eeqn
The arguments of the left-hand side label the momenta of the propagators in 
successive order.  Note that the scalar integrals are invariant under
inversion of the order of the momenta; {\it e.g.} 
$D_{0}(p;k;v)=D_{0}(v;k;p)$.
The integrals $C_{0}$ and $D_{0}$ are expressly finite in four 
dimensions, while the integral $B_{0}$ only occurs in combinations 
that are finite as $\epsilon\equiv (4-n)/2\rightarrow0$.  Therefore, we have 
explicitly removed the $1/\epsilon$-pole from its definition.

Explicit formulae for some of the scalar integrals are
\bea
\eqalign
{\vcenter{\openup2\jot\halign{
    $\displaystyle\hfil#$&$\displaystyle#\hfil$\cr
   B_0(q) =\ & -{i\over8\pi^2}\sqrt{{4\mt^2-q^2\over q^2}}\tan^{-1}
       \sqrt{q^2\over4\mt^2-q^2}\qquad {\rm if}\ 0 < q^2 < 4\mt^2,\cr
   B_0(q) =\ & -{i\over16\pi^2}\sqrt{{q^2-4\mt^2\over q^2}}
       \ln{1+\sqrt{q^2\over q^2-4\mt^2}\over1-\sqrt{q^2
       \over q^2-4\mt^2}}\qquad
       {\rm if}\ q^2 < 0\ {\rm or}\ q^2 > 4\mt^2,\cr
   C_0(q_{1};q_{2}) =\ & {i\over16\pi^2}{1\over\sqrt{\Delta_3}}
      \bigg\{\ln(1-y_-)\ln\left({1-y_-\de_1^+
            \over1-y_-\de_1^-}\right)\cr
      &+\ \ln(1-x_-)\ln\left({1-x_-\de_2^+
            \over1-x_-\de_2^-}\right)
       +\ \ln(1-z_-)\ln\left({1-z_-\de_3^+
            \over1-z_-\de_3^-}\right)\cr
      &+\ {\rm Li}_2\left({y_+\de_1^+}\right)
       +\ {\rm Li}_2\left({y_-\de_1^+}\right)
       -\ {\rm Li}_2\left({y_+\de_1^-}\right)
       -\ {\rm Li}_2\left({y_-\de_1^-}\right)\cr
      &+\ {\rm Li}_2\left({x_+\de_2^+}\right)
       +\ {\rm Li}_2\left({x_-\de_2^+}\right)
       -\ {\rm Li}_2\left({x_+\de_2^-}\right)
       -\ {\rm Li}_2\left({x_-\de_2^-}\right)\cr
      &+\ {\rm Li}_2\left({z_+\de_3^+}\right)
       +\ {\rm Li}_2\left({z_-\de_3^+}\right)
       -\ {\rm Li}_2\left({z_+\de_3^-}\right)
       -\ {\rm Li}_2\left({z_-\de_3^-}\right)\bigg\}\, ,\cr
}}}
\label{finiteints}
\eea
where 
\bea
Q&=&-q_{1}-q_{2}\nonumber
\eea
and
\bea
\Delta_3 &=& (q_1^2)^2 + (q_2^2)^2 + (Q^2)^2 - 2q_1^2q_2^2 -2q_1^2Q^2
- 2q_2^2Q^2 \nonumber
\eea
\beq
   \delta_1= {q_1^2-q_2^2-Q^2\over\sqrt{\Delta_3}}\ ,\qquad 
       \delta_2 = {-q_1^2+q_2^2-Q^2\over\sqrt{\Delta_3}}\ ,\qquad
       \delta_3 = {-q_1^2-q_2^2+Q^2\over\sqrt{\Delta_3}}\ ,\label{littledeltas}
\eeq
\beq
\de_1^+ = \frac{1+\de_1}{2}, \quad \de_2^+ = \frac{1+\de_2}{2},
\quad  \de_3^+ = \frac{1+\de_3}{2},
\eeq
\beq
\de_1^- = \frac{1-\de_1}{2}, \quad \de_2^- = \frac{1-\de_2}{2},
\quad  \de_3^- = \frac{1-\de_3}{2},
\eeq
\bea
   x_\pm &=& {q_2^2\over2\mt^2} \pm
          {q_2^2\over2\mt^2}\sqrt{1-{4\mt^2\over q_2^2}}\ ,\quad
      y_\pm = {q_1^2\over2\mt^2} \pm
          {q_1^2\over2\mt^2}\sqrt{1-{4\mt^2\over q_1^2}}\ ,\nn\\
      z_\pm &=& {Q^2\over2\mt^2} \pm
          {Q^2\over2\mt^2}\sqrt{1-{4\mt^2\over Q^2}}
\eea

\section{The top triangle with two off-shell gluons}
\label{sec:triangleloop}

The triangle vertex for two off-shell gluons of momenta $q_1^\mu$ and
$q_2^\nu$ and color indexes $a$ and $b$ respectively, and a Higgs boson of
momentum $Q$, with $q_1+q_2+Q=0$ is given by
\beq
  {\cal A}^{\mu\nu} = 4\,\delta^{ab}\,{g^2\mt^2\over\ v}\Bigl[
       q_2^\mu q_1^\nu A_1(q_1;q_2) - g^{\mu\nu} A_2(q_1;q_2) \Bigr]\, ,
\label{Dj}
\eeq
where we have dropped all terms proportional to $q_1^\mu$ and $q_2^\nu$,
since they are contracted with external conserved currents, and 
with
\bea
\eqalign
{\vcenter{\openup2\jot\halign{
    $\displaystyle\hfil#$&$\displaystyle#\hfil$\cr
A_1(q_1;q_2) = &\
       C_0(q_{1};q_{2})\left[{4\mt^2\over\Delta_3}\(Q^2-q_1^2-q_2^2\)-1 -
       {4q_1^2q_2^2\over\Delta_3} - {12q_1^2q_2^2Q^2\over\Delta_3^2}
       \left(q_1^2+q_2^2-Q^2\right)\right] \cr
     &       -\lq B_0(q_2)-B_{0}(Q)\rq\left[{2q_2^2\over\Delta_3}
       + {12q_1^2q_2^2\over\Delta_3^2}
         \left(q_2^2-q_1^2+Q^2\right)\right]\cr
     &
       -\lq B_0(q_1)-B_{0}(Q)\rq \left[{2q_1^2\over\Delta_3}
       +{12q_1^2q_2^2\over\Delta_3^2}\left(q_1^2-q_2^2+Q^2
       \right)
       \right] \cr
     &
       - {2\over\Delta_3} {i\over (4\pi)^2} \left(q_1^2+q_2^2-Q^2\right)\cr
A_2(q_1;q_2) = &\
C_0(q_{1};q_{2})\left[2\mt^2 + {1\over2}\left(
        q_1^2+q_2^2-Q^2\right)
        +{2q_1^2q_2^2Q^2\over\Delta_3}\right]\cr
      &       
       +\lq B_0(q_2)-B_{0}(Q)\rq {1\over\Delta_3} q_2^2 
      \left(q_2^2-q_1^2-Q^2\right)\cr  
      &+ \lq B_0(q_1)-B_{0}(Q)\rq {1\over\Delta_3}
        q_1^2\left(q_1^2-q_2^2-Q^2\right)+{i\over (4\pi)^2} \cr
}}}
\label{a1a2}
\eea
with the scalar integrals evaluated in Appendix~\ref{sec:scalars}.
In the infinite top-mass limit~(\ref{efflag}), 
the coefficients $A_{1,2}$ become
\beq
\label{a1a2t} 
\lim_{\mt\to\infty} 4\,{g^2\mt^2\over\ v} \, A_1 = i A \qquad\qquad
\lim_{\mt\to\infty} 4\,{g^2\mt^2\over\ v} \, A_2 = i\, q_1\cdot q_2 A\,.
\eeq
where $A$ is given in Eq.~(\ref{eq:A_def}).

\section{The top box diagram with one off-shell gluon}
\label{sec:boxloop}

There are three distinct box diagrams (including both directions of fermion
flow) with two on-shell gluons with momenta $k_1^\mu$ and $k_2^\nu$ and color
indexes $a$ and $b$, one off-shell gluon with momentum $k^\rho\equiv
k_3^\rho+k_4^\rho$ and color index $c$, and a Higgs boson of momentum $k_H$,
with $k_1+k_2+k+k_H=0$.  The sum of the three box diagrams
can be parametrized
in terms of 14 form factors $H_i$ by
\beq
\cM^{\mu\nu\rho}\ =\  {g^3\mt^2\over v}\, 2\,i f^{bac}\, J^{\mu\nu\rho}
\eeq
with
\bea
J^{\mu\nu\rho} &=&
g^{\mu\nu} \left( H_1\ k_1^\rho + H_2\ k_2^\rho \right)
+ g^{\mu\rho} \left( H_3\ k_1^\nu + H_4\ k^\nu \right) +
g^{\nu\rho} \left( H_5\ k^\mu + H_6\ k_2^\mu \right) \nonumber\\ &+&
H_7\ k_1^\nu k_2^\mu k_2^\rho + H_8\ k^\nu k_2^\mu k_2^\rho +
H_9\ k_2^\mu k_1^\nu k_1^\rho + H_{10}\ k_2^\rho k^\mu k^\nu \label{boxfin}\\ 
&+& H_{11}\ k^\mu k_1^\nu k_1^\rho + H_{12}\ k_1^\rho k^\mu k^\nu +
H_{13}\ k_2^\mu k_1^\rho k^\nu + H_{14}\ k^\mu k_1^\nu k_2^\rho\ , \nonumber
\eea
where we have  removed terms with $k_1^\mu$, $k_2^\nu$, and $k^\rho$,
using the fact that the contraction with the respective polarization vector
is zero ($k_i\cdot \epsilon(k_i)=0$) and that the off-shell gluon 
is going to be contracted with an on-shell fermion or gluon current.

As discussed in Section~\ref{sec:hqqgg}, we can express the amplitudes for 
$q\,g\to q\,g\,H$ scattering in terms of just 6 of the form factors:
$H_1, H_2, H_4,H_5,H_{10},H_{12}$.  In addition, these are related by
\bea
H_2&=&-H_1  \{k_1\leftrightarrow k_2\}\nonumber\\
H_5&=&-H_4  \{k_1\leftrightarrow k_2\}\\
H_{12}&=&-H_{10}  \{k_1\leftrightarrow k_2\}\ .\nonumber
\eea
In the interest of brevity we will therefore only give the expressions for
$H_1,H_4,H_{10}$.  

We write the form factors as the sum of the contributions from the three
distinct gluon orderings in the box diagram, $H_i=E_i+F_i+G_i$.  They are
functions of the scalar integrals, $B_{0}(q)$, $C_{0}(q;p)$, and
$D_{0}(q;p;k)$, defined in Appendix~\ref{sec:scalars}, and are related
to the box form factors of Appendix~D of Ref.~\cite{DelDuca:2001fn}.
For notational
convenience we write the arguments of the scalar integrals by the number of
the parton momentum, combining numbers if the momenta are added: {\it e.g.}
$C_{0}(12;34)\equiv C_{0}(k_{1}+k_{2};\,k_{3}+k_4)$.  The coefficients of the
scalar integrals are functions of four variables $s_{12}, s_{34},S_1,S_2$
with
\bea
s_{ij}&=&(k_i+k_j)^2\nonumber\\
S_{i}&=& 2\,k_i\cdot k\ =\ s_{i3}+s_{i4}\label{detiii}\ .
\eea
To simplify the expressions we also find it useful to introduce the following
variables, which arise as determinants in the Passarino-Veltman
reduction~\cite{Passarino:1979jh},
\bea
\Delta&=& s_{12}s_{34}-S_1S_2\label{detii}\\
\Sigma&=& 4s_{12}s_{34}-(S_1+S_2)^2\ .
\nonumber
\eea
In terms of these variables we obtain
{\small
\bea
E_1&=&  -s_{12}D_0(2;1;34)\Biggl[1-\frac{8\mt^2}{s_{12}}+\frac{S_2}{2s_{12}}+
\frac{S_2(s_{12}-8\mt^2)(s_{34}+S_1)}{2s_{12}\Delta}\nonumber\\
&&\qquad\qquad
+\frac{2(s_{34}+S_1)^2}{\Delta}+\frac{S_2(s_{34}+S_1)^3}{\Delta^2}
\Biggr]\nonumber\\
&&-\Bigl[\left(s_{12}+S_2\right)C_0(2;134)
-s_{12}C_0(1;2)+\left(S_1-S_2\right)C_0(12;34)
-S_1C_0(1;34)\Bigr]\nonumber\\
&&\qquad\qquad\times\Biggl(\frac{S_2(s_{12}-4\mt^2)}{2s_{12}\Delta}+
\frac{2(s_{34}+S_1)}{\Delta}+\frac{S_2(s_{34}+S_1)^2}
{\Delta^2}\Biggr)\nonumber\\
&&+\left[C_0(1;34)-C_0(12;34)\right]\left(1-\frac{4\mt^2}{s_{12}}\right)
-C_0(12;34)\,\frac{2s_{34}}{S_1}
\\
&&-\left[B_0(134)-B_0(1234)\right]\,\frac{2s_{34}(s_{34}+S_1)}{S_1\Delta}\nonumber\\
&&+\left[B_0(34)-B_0(1234)+s_{12}C_0(12;34)\right]\,\left(\frac{2
s_{34}(s_{34}+S_1)(S_1-S_2)}{\Delta\Sigma}+\frac{2s_{34}(s_{34}+S_1)}{S_1\Delta}
\right)\nonumber\\
&&+\left[B_0(12)
-B_0(1234)-(s_{34}+S_1+S_2)C_0(12;34)\right]
\,\frac{2(s_{34}+S_1)(2s_{12}s_{34}-S_2(S_1+S_2))}{\Delta\Sigma}\nonumber\\
F_1&=&  -S_{2}D_0(1;2;34)\Biggl[\frac{1}{2}-
\frac{(s_{12}-8\mt^2)(s_{34}+S_2)}{2\Delta}
-\frac{s_{12}(s_{34}+S_2)^3}{\Delta^2}
\Biggr]\nonumber\\
&&+\Bigl[\left(s_{12}+S_1\right)C_0(1;234)
-s_{12}C_0(1;2)-\left(S_1-S_2\right)C_0(12;34)
-S_2C_0(2;34)\Bigr]\nonumber\\
&&\qquad\qquad\times\Biggl(\frac{S_2(s_{12}-4\mt^2)}{2s_{12}\Delta}
+\frac{S_2(s_{34}+S_2)^2}
{\Delta^2}\Biggr)\nonumber\\
&&-\left[C_0(12;34)-C_0(1;234)\right]\left(1-\frac{4\mt^2}{s_{12}}\right)
-C_0(1;234)
\nonumber\\
&&+\left[B_0(234)-B_0(1234)\right]\,\frac{2(s_{34}+S_2)^2}{(s_{12}+S_1)\Delta}\\
&&-\left[B_0(34)-B_0(1234)+s_{12}C_0(12;34)\right]\,\frac{2
s_{34}(s_{34}+S_2)(S_2-S_1)}{\Delta\Sigma}\nonumber\\
&&+\left[B_{0}(12)
-B_0(1234)-(s_{34}+S_1+S_2)C_0(12;34)\right]
\,\frac{2(s_{34}+S_2)(2s_{12}s_{34}-S_2(S_1+S_2))}{\Delta\Sigma}\nonumber\\
G_1&=&  S_{2}D_0(1;34;2)\Biggl[\frac{\Delta}{s_{12}^2}
-\frac{4\mt^2}{s_{12}}
\Biggr]\nonumber\\
&&-S_2\Bigl[\left(s_{12}+S_1\right)C_0(1;234)
-S_1C_0(1;34)\Bigr]\left(\frac{1}{s_{12}^2}-\frac{s_{12}-4\mt^2}{2s_{12}\Delta}\right)\nonumber\\
&&-S_2\Bigl[\left(s_{12}+S_2\right)C_0(13;2)
-S_2C_0(2;34)\Bigr]\left(\frac{1}{s_{12}^2}+\frac{s_{12}-4\mt^2}{2s_{12}\Delta}\right)\nonumber\\
&&-C_0(1;34)-[C_0(1;234)-C_0(1;34)]\frac{4\mt^2}{s_{12}}
+\left[B_0(134)-B_0(1234)\right]\,\frac{2}{s_{12}}\\
&&+\left[B_0(134)-B_0(34)\right]\,\frac{2s_{34}}{s_{12}S_1}+\left[B_0(234)-B_0(1234)\right]
\,\frac{2(s_{34}+S_2)}{s_{12}(s_{12}+S_1)}\nonumber\\
E_4&=&  -s_{12}D_0(2;1;34)\Biggl[\frac{1}{2}-
\frac{(S_1-8\mt^2)(s_{34}+S_1)}{2\Delta}
-\frac{s_{12}(s_{34}+S_1)^3}{\Delta^2}
\Biggr]\nonumber\\
&&+\Bigl[\left(s_{12}+S_2\right)C_0(2;134)
-s_{12}C_0(1;2)+\left(S_1-S_2\right)C_0(12;34)
-S_1C_0(1;34)\Bigr]\nonumber\\
&&\qquad\qquad\times\Biggl(\frac{(S_1-4\mt^2)}{2\Delta}
+\frac{s_{12}(s_{34}+S_1)^2}
{\Delta^2}\Biggr)\\
&&-C_0(12;34)+\left[B_0(134)-B_0(1234)\right]\,\left(\frac{2s_{34}}{\Delta}+
\frac{2s_{12}(s_{34}+S_1)}{(s_{12}+S_2)\Delta}\right)
\nonumber\\
&&-\left[B_0(34)-B_0(1234)+s_{12}C_0(12;34)\right]\,\left(\frac{2
s_{34}(2s_{12}s_{34}-S_2(S_1+S_2)+s_{12}(S_1-S_2))}{\Delta\Sigma}
\right)\nonumber\\
&&+\left[B_0(12)
-B_0(1234)-(s_{34}+S_1+S_2)C_0(12;34)\right]\nonumber\\
&&\qquad\qquad\times
\,\left(\frac{2s_{12}(2s_{12}s_{34}-S_1(S_1+S_2)+s_{34}(S_2-S_1))}
{\Delta\Sigma}\right)\nonumber\\
F_4&=&  -s_{12}D_0(1;2;34)\Biggl[\frac{1}{2}+
\frac{(S_1-8\mt^2)(s_{34}+S_2)}{2\Delta}
+\frac{s_{12}(s_{34}+S_2)^3}{\Delta^2}
\Biggr]\nonumber\\
&&-\Bigl[\left(s_{12}+S_1\right)C_0(1;234)
-s_{12}C_0(1;2)-\left(S_1-S_2\right)C_0(12;34)
-S_2C_0(2;34)\Bigr]\nonumber\\
&&\qquad\qquad\times\Biggl(\frac{(S_1-4\mt^2)}{2\Delta}
+\frac{s_{12}(s_{34}+S_2)^2}
{\Delta^2}\Biggr)\\
&&-C_0(12;34)-\left[B_0(234)-B_0(1234)\right]\,\frac{2(s_{34}+S_2)}{\Delta}\nonumber\\
&&+\left[B_0(34)-B_0(1234)+s_{12}C_0(12;34)\right]\,\frac{2
s_{34}(2s_{12}s_{34}-S_1(S_1+S_2)+s_{12}(S_2-S_1))}{\Delta\Sigma}\nonumber\\
&&-\left[B_{0}(12)
-B_0(1234)-(s_{34}+S_1+S_2)C_0(12;34)\right]\nonumber\\
&&\qquad\qquad\times
\,\left(\frac{2s_{12}(2s_{12}s_{34}-S_2(S_1+S_2)+s_{34}(S_1-S_2))}
{\Delta\Sigma}\right)\nonumber\\
G_4&=&  -D_0(1;34;2)\Biggl[\frac{\Delta}{s_{12}}+\frac{s_{12}+S_1}{2}
-4\mt^2
\Biggr]\nonumber\\
&&+\Bigl[\left(s_{12}+S_1\right)C_0(1;234)
-S_1C_0(1;34)\Bigr]\left(\frac{1}{s_{12}}-\frac{S_1-4\mt^2}{2\Delta}\right)\nonumber\\
&&+\Bigl[\left(s_{12}+S_2\right)C_0(13;2)
-S_2C_0(2;34)\Bigr]\left(\frac{1}{s_{12}}+\frac{S_1-4\mt^2}{2\Delta}\right)\\
&&+\left[B_0(1234)-B_0(134)\right]\,\frac{2}{s_{12}+S_2}\nonumber\\
E_{10}&=&  -s_{12}D_0(2;1;34)\Biggl[\frac{s_{34}+S_1}{\Delta}+
\frac{12\mt^2S_1(s_{34}+S_1)}{\Delta^2}
-\frac{4s_{12}S_1(s_{34}+S_1)^3}{\Delta^3}
\Biggr]\nonumber\\
&&-\Bigl[\left(s_{12}+S_2\right)C_0(2;134)
-s_{12}C_0(1;2)+\left(S_1-S_2\right)C_0(12;34)
-S_1C_0(1;34)\Bigr]\nonumber\\
&&\qquad\qquad\times\Biggl(\frac{1}{\Delta}+\frac{4\mt^2S_1}{\Delta^2}
-\frac{4s_{12}S_1(s_{34}+S_1)^2}
{\Delta^3}\Biggr)\nonumber\\
&&+C_0(12;34)\left(\frac{4s_{12}s_{34}(S_1-S_2)}{\Delta\Sigma}
-\frac{4(s_{12}-2\mt^2)(2s_{12}s_{34}-S_1(S_1+S_2))}{\Delta\Sigma}
\right)\\
&&+\left[B_0(134)-B_0(1234)\right]\,\left(
\frac{4(s_{34}+S_1)}{(s_{12}+S_2)\Delta}+
\frac{8S_1(s_{34}+S_1)}{\Delta^2}\right)
\nonumber\\
&&+\left[B_0(34)-B_0(1234)+s_{12}C_0(12;34)\right]\,\Biggl(
\frac{12s_{34}(2s_{12}+S_1+S_2)(2s_{12}s_{34}-S_1(S_1+S_2))}{\Delta\Sigma^2}
\nonumber\\
&&\qquad\qquad-\frac{4s_{34}(4s_{12}+3S_1+S_2)}{\Delta\Sigma}
+\frac{8s_{12}s_{34}(s_{34}(s_{12}+S_2)-S_1(s_{34}+S_1))}{\Delta^2\Sigma}
\Biggr)\nonumber\\
&&+\left[B_0(12)
-B_0(1234)-(s_{34}+S_1+S_2)C_0(12;34)\right]\nonumber\\
&&\qquad\qquad\times
\,\Biggl(\frac{12s_{12}(2s_{34}+S_1+S_2)
(2s_{12}s_{34}-S_1(S_1+S_2))}{\Delta\Sigma^2}\nonumber\\
&&\qquad\qquad\qquad
+\frac{8s_{12}S_1(s_{34}(s_{12}+S_2)-S_1(s_{34}+S_1))}{\Delta^2\Sigma}\Biggr)\nonumber\\
&&+\left(\frac{i}{4\pi^2}\right)\left(\frac{2s_{12}s_{34}-S_1(S_1+S_2)}{\Delta\Sigma}\right)
\nonumber\\
F_{10}&=&  s_{12}D_0(1;2;34)\Biggl[\frac{s_{34}+S_2}{\Delta}-\frac{4\mt^2}{\Delta}
+\frac{12\mt^2s_{34}(s_{12}+S_1)}{\Delta^2}\nonumber\\
&&\qquad\qquad-\frac{4s_{12}(s_{34}+S_2)^2}{\Delta^2}
-\frac{4s_{12}S_1(s_{34}+S_2)^3}{\Delta^3}
\Biggr]\nonumber\\
&&+\Bigl[\left(s_{12}+S_1\right)C_0(1;234)
-s_{12}C_0(1;2)-\left(S_1-S_2\right)C_0(12;34)
-S_2C_0(2;34)\Bigr]\nonumber\\
&&\qquad\qquad\times\Biggl(\frac{1}{\Delta}+\frac{4\mt^2S_1}{\Delta^2}
-\frac{4s_{12}(s_{34}+S_2)}{\Delta^2}
-\frac{4s_{12}S_1(s_{34}+S_2)^2}{\Delta^3}\Biggr)\nonumber\\
&&-C_0(12;34)\left(\frac{4s_{12}s_{34}}{S_2\Delta}
+\frac{4s_{12}s_{34}(S_2-S_1)}{\Delta\Sigma}
+\frac{4(s_{12}-2\mt^2)(2s_{12}s_{34}-S_1(S_1+S_2))}{\Delta\Sigma}
\right)\nonumber\\
&&-\left[B_0(234)-B_0(1234)\right]\,\left(
\frac{4s_{34}}{S_2\Delta}+
\frac{8s_{34}(s_{12}+S_1)}{\Delta^2}\right)\\
&&-\left[B_0(34)-B_0(1234)+s_{12}C_0(12;34)\right]\,\Biggl(
-\frac{12s_{34}(2s_{12}+S_1+S_2)(2s_{12}s_{34}-S_1(S_1+S_2))}{\Delta\Sigma^2}
\nonumber\\
&&\qquad\qquad-\frac{4s_{12}s_{34}^2}{S_2\Delta^2}+\frac{4s_{34}S_1}{\Delta\Sigma}
-\frac{4s_{34}(s_{12}s_{34}(2s_{12}+S_2)-S_1^2(2s_{12}+S_1))}{\Delta^2\Sigma}
\Biggr)\nonumber\\
&&-\left[B_{0}(12)
-B_0(1234)-(s_{34}+S_1+S_2)C_0(12;34)\right]\nonumber\\
&&\qquad\qquad\times
\,\Biggl(-\frac{12s_{12}(2s_{34}+S_1+S_2)
(2s_{12}s_{34}-S_1(S_1+S_2))}{\Delta\Sigma^2}
+\frac{8s_{12}(2s_{34}+S_1)}{\Delta\Sigma}\nonumber\\
&&\qquad\qquad
-\frac{8s_{12}s_{34}(2s_{12}s_{34}-S_1(S_1+S_2)+s_{12}(S_2-S_1))}
{\Delta^2\Sigma}\Biggr)\nonumber\\
&&+\left(\frac{i}{4\pi^2}\right)\left(\frac{2s_{12}s_{34}-S_1(S_1+S_2)}{\Delta\Sigma}\right)
\nonumber\\
G_{10}&=&  -D_0(1;34;2)\left(1+\frac{4S_1\mt^2}{\Delta}\right)\nonumber\\
&&+\Bigl[\left(s_{12}+S_1\right)C_0(1;234)
-S_1C_0(1;34)\Bigr]\left(\frac{1}{\Delta}+\frac{4S_1\mt^2}{\Delta^2}\right)\nonumber\\
&&-\Bigl[\left(s_{12}+S_2\right)C_0(13;2)
-S_2C_0(2;34)\Bigr]\left(\frac{1}{\Delta}+\frac{4S_1\mt^2}{\Delta^2}\right)\\
&&+\left[B_0(1234)-B_0(134)\right]\,\frac{4(s_{34}+S_1)}{\Delta(s_{12}+S_2)}
+\left[B_0(34)-B_0(234)\right]\,\frac{4s_{34}}{\Delta S_2}\nonumber
\eea
}

\section{Kinematics for Higgs  + 2 jets}
\label{sec:appa}

We consider the production of two partons of momentum $p_1$ and $p_3$
and a Higgs boson of momentum $\ph$,
in the scattering between two partons of momenta $p_2$ and $p_4$, where 
all momenta are taken as outgoing.
Using light-cone coordinates $p^{\pm}= p_0\pm p_z $, and
complex transverse coordinates $p_{\perp} = p_x + i p_y$, with scalar
product $2\, p\cdot q = p^+q^- + p^-q^+ - p_{\perp} q^*_{\perp} - p^*_{\perp} 
q_{\perp}$, the 4-momenta are
\begin{eqnarray}
p_2 &=& \left(p_2^+/2, 0, 0,p_2^+/2 \right) 
     \equiv  \left(p_2^+ , 0; 0, 0 \right)\, ,\nonumber \\
p_4 &=& \left(p_4^-/2, 0, 0,-p_4^-/2 \right) 
     \equiv  \left(0, p_4^-; 0, 0\right)\, ,\label{in}\\
p_i &=& \left( (p_i^+ + p_i^- )/2, 
                {\rm Re}[\pip],
                {\rm Im}[\pip], 
                (p_i^+ - p_i^- )/2 \right)\nonumber\\
    &\equiv& \left(\mip e^{y_i}, \mip e^{-y_i}; 
|\pip|\cos{\phi_i}, |\pip|\sin{\phi_i}\right) \quad
i = 1, 3, H\, ,\nonumber
\end{eqnarray}
where $y$ is the rapidity and $m_{\sss \perp} = \sqrt{p_{\sss \perp}^2+m^2}$ 
the transverse mass, which for massless particles reduces to 
$|p_{\sss\perp}|$. 
The first notation in \eqn{in} is the standard representation 
$p^\mu =(p_0,p_x,p_y,p_z)$, while the second features light-cone
components, on which we have used the mass-shell condition.
Momentum conservation implies
\begin{eqnarray}
0 &=& p_{1_\perp} + p_{3_\perp} + \php\, ,\nonumber \\
p_2^+ &=& - p_1^+ - p_3^+ - \ph^+\, ,\label{nkin}\\ 
p_4^- &=& - p_1^- - p_3^- - \ph^-\, .\nonumber
\end{eqnarray}

Using the spinor representation of Ref.~\cite{DelDuca:1995zy,DelDuca:2000ha},
the spinor products~(\ref{spino}) are
\begin{eqnarray}
\langle p_i p_j\rangle &=& \pip\sqrt{p_j^+\over p_i^+} - \pjp
\sqrt{p_i^+\over p_j^+}\, , \nonumber\\ 
\langle p_2 p_i\rangle &=& - i \sqrt{-p_2^+
\over p_i^+}\, \pip\, ,\label{spro}\\ 
\langle p_i p_4\rangle &=&
i \sqrt{-p_4^- p_i^+}\, ,\nonumber\\ 
\langle p_2 p_4\rangle 
&=& -\sqrt{s}\, ,\nonumber
\end{eqnarray}
with $i = 1, 3$, and 
where we have used the mass-shell condition 
$|\pip|^2 = p_i^+ p_i^-$. The currents are obtained from \eqn{currentsi}.

In the high-energy limit~(\ref{mrk}), the $\pm$ components of the 
momentum conservation~(\ref{nkin}) become
\begin{eqnarray}
p_2^+ &\simeq& -p_1^+\, ,\label{mrkin}\\ 
p_4^- &\simeq& -p_3^-\, .\nonumber
\end{eqnarray}
In the high-energy limit~(\ref{mrk2}), we have instead
\begin{eqnarray}
p_2^+ &\simeq& -p_1^+ - \ph^+\, ,\label{mrkin2}\\ 
p_4^- &\simeq& -p_3^-\, .\nonumber
\end{eqnarray}
A full list of the spinor products~(\ref{spro}) in the limits~(\ref{mrkin})
and~(\ref{mrkin2}) can be found in Appendices B and C of 
Ref.~\cite{DelDuca:2000ha}, and will not be repeated here.

%
%

\section{Effective vertices for the production of a Higgs boson plus one jet}
\label{sec:vertex2}

We start from  the sub-amplitude~(\ref{apmii}) for $q\, g\to q\, g\, H$
scattering. 
In order to make high-energy expansions simpler, we rewrite it as
\bea
\lefteqn{i\ m(1^+, 2^-; 3_q^+, 4_{\qb}^-) = - 
{g^4\mt^2\over\ v}\,\frac{1}{s_{12}s_{34}}\ \frac{1}{\langle13\rangle[24]} }
\nonumber\\ && \Biggl\{ \frac{1}{2}
\biggl( \langle 2|3|1\rangle + \langle 2|4|1\rangle\biggr)\biggl[
4s_{24}s_{13}\ (H_4+H_5) \nonumber\\
&&\qquad -s_{13}\Delta H_{12}-s_{24}\Delta H_{10}
+2(s_{23}s_{13}-s_{24}s_{14})\ (s_{24} H_{10} -s_{13} H_{12})\nonumber\\
&&\qquad+4s_{24}(s_{23}-s_{24})A_1(2; 134)
+4s_{13}(s_{13}-s_{14})A_1(1; 234)\biggr]\nonumber\\
&&
+\frac{1}{2}
\biggl( \langle 2|3|1\rangle - \langle 2|4|1\rangle\biggr)\biggl[
\Delta\ (s_{24} H_{10} -s_{13} H_{12})\nonumber\\
&&\qquad-4s_{24}(s_{23}+s_{24})A_1(2; 134)
-4s_{13}(s_{13}+s_{14})A_1(1; 234)\biggr]\Biggr\}
\, .\label{apmii2}
\eea
Then we take the partons 2 and 4 as the incoming gluon and quark,
respectively, and consider the high-energy limit~(\ref{mrk2})
\beq
p_1^+, \ph^+ \gg p_3^+ \qquad p_1^-, \ph^- \ll p_3^-\ .
\eeq
In this high energy limit the matrix element 
should rise as ${\cal O}(s)$.  Thus, we need
to find the high energy behavior of every quantity in~(\ref{apmii2}).
We note that $\langle13\rangle[24]$ is ${\cal O}(s)$. In addition,
the currents that appear in \eqn{apmii2} are
\bea
\langle 2|3|1\rangle - \langle 2|4|1\rangle&=& i \sqrt{-p_2^+ \over p_1^+} 
( 2\,p_1^+ p_3^-)\ +\ {\cal O}(s^0)\nonumber\\
\langle 2|3|1\rangle + \langle 2|4|1\rangle &=& -i \sqrt{-p_2^+ \over
p_1^+} (p_{1\perp}^* p_{3\perp} + |p_{1\perp}|^2 + p_1^+ \ph^-)
\ ,\label{currents}
\eea
with $p_2^+ \simeq -(p_1^+ + \ph^+)$. Note that the current 
$(\langle 2|3|1\rangle + \langle 2|4|1\rangle)$ is exact 
and ${\cal O}(s^0)$, thus we need only keep the terms that are ${\cal O}(s^2)$
in the brackets of \eqn{apmii2} that multiply it. Since the current
$(\langle 2|3|1\rangle - \langle 2|4|1\rangle)$
is ${\cal O}(s)$, we need only keep the terms that are ${\cal O}(s)$
in the brackets of \eqn{apmii2} that multiply it.  It is easy to see that the 
${\cal O}(s^2)$ terms in the second brackets will vanish.

The $H$ and $A$ functions all scale as ${\cal O}(s^0)$.  So we only
need to get the scaling behavior of the six $s_{ij}$ invariants.
There are just two distinct invariants that rise as ${\cal O}(s)$, which
can be taken to be $s_{13}$ and $s_{24}$.  We then define
\begin{eqnarray}
S_1&=&s_{13}+s_{14}\nonumber\\
S_2&=&s_{23}+s_{24}\ .
\label{si}
\end{eqnarray}
These invariants are ${\cal O}(s^0)$.
Then we can replace $s_{14}=-s_{13}+S_1$ and $s_{23}=-s_{24}+S_2$
everywhere, leaving us with two invariants of ${\cal O}(s)$,
$s_{13}$ and $s_{24}$, and four invariants of ${\cal O}(s^0)$,
$s_{12}$, $s_{34}$, $S_1$, and $S_2$. Note that these last four invariants 
are just those that arise in the $H$ and $A$ functions. In addition,
note that the quantity $\Delta$, \eqn{deti}, is ${\cal O}(s^0)$, since it is
$\Delta\ =\ s_{12}s_{34}-S_1S_2$.

Then the high-energy expansion of \eqn{apmii2} is
\bea
\lefteqn{i\ m(1^+, 2^-; 3_q^+, 4_{\qb}^-) = - 
{g^4\mt^2\over\ v}\,\frac{1}{s_{12}s_{34}}\ \frac{1}{\langle13\rangle[24]} }
\nonumber\\ && \Biggl\{ \frac{1}{2}
\biggl( \langle 2|3|1\rangle + \langle 2|4|1\rangle\biggr)\biggl[
(s_{13})^2\biggl(8A_1(1; 234)-2S_2H_{12}\biggr)\nonumber\\
&&\qquad +(s_{24})^2\biggl(-8A_1(2; 134)-2S_1H_{10}\biggr)\nonumber\\
&&\qquad +s_{13}s_{24}\biggl(4(H_4+H_5)+2S_2H_{10}+2S_1H_{12}\biggr)
\biggr]\nonumber\\
&& + \frac{1}{2}
\biggl( \langle 2|3|1\rangle - \langle 2|4|1\rangle\biggr)\biggl[
s_{13}\biggl(-\Delta H_{12}-4S_1A_1(1; 234)\biggr)\nonumber\\
&&\qquad+s_{24}\biggl(\Delta H_{10}-4S_2A_1(2; 134)\biggr)
\biggr]\Biggr\}\ +\ {\cal O}(s^0)\, ,\label{apmiii}
\eea
where the invariants are
\bea
s_{13}&=&p_1^+p_3^-\ +\ {\cal O}(s^0)\nonumber\\
s_{24}&=&(p_1^+ +\ph^+)p_3^-\ +\ {\cal O}(s^0)\nonumber\\
s_{12}&=&-(|p_{1\perp}|^2+\ph^+p_1^-)\ +\ {\cal O}(s^{-1})\nonumber\\
s_{34}&=&-|p_{3\perp}|^2\ +\ {\cal O}(s^{-1})\nonumber\\
S_1&=&|p_{3\perp}|^2-|\php|^2-p_1^+\ph^-\ +\ {\cal O}(s^{-1})\nonumber\\
S_2&=&|p_{1\perp}|^2+\mhp^2 + p_1^+ \ph^- + \ph^+ p_1^-
\ +\ {\cal O}(s^{-1})\ .\label{invariants}
\eea

In order to analyze the sub-amplitude~(\ref{appii}), we need the currents
\bea
\langle 4|1|3\rangle &=& i \sqrt{-p_4^-\over p_3^-} 
{p_{3\perp}^*\over |p_{3\perp}|} p_1^+ p_3^- + {\cal O}(s^0)
\nonumber\\
\langle 4|2|3\rangle &=& - i \sqrt{-p_4^-\over p_3^-} 
{p_{3\perp}^*\over |p_{3\perp}|} (p_1^+ + \ph^+) p_3^- + 
{\cal O}(s^0)\, . \label{curr2}
\eea
The sum and the difference of the currents~(\ref{curr2}) are still 
${\cal O}(s)$. Rather, it is convenient to consider the linear combination
\bea
\langle 4|1|3\rangle s_{24} + \langle 4|2|3\rangle s_{13} 
&=& i \sqrt{-p_4^-\over p_3^-} {p_{3\perp}^*\over |p_{3\perp}|} 
\Bigl[ (p_1^+ + \ph^+) p_3^- p_{1\perp} p_{3\perp}^* 
\nonumber\\ && \quad
+\ p_1^+ p_3^- ( |p_{1\perp}|^2 + \mhp^2 + p_1^+ \ph^- + \ph^+ p_1^-)
\Bigr] + {\cal O}(s^0)\ ,\label{curr3}
\eea
which is also ${\cal O}(s)$ (all the ${\cal O}(s^2)$ terms cancel out of it).
Expanding the invariants in terms of $s_{13}$ and $s_{24}$, using
\eqn{si}, and collecting all the ${\cal O}(s)$ terms,
we obtain the high-energy expansion of \eqn{appii}
\bea
i\ m(1^+, 2^+; 3_q^+, 4_{\qb}^-) &=& - 
{g^4\mt^2\over\ v}\,\frac{1}{s_{12} s_{34}}\ {[21]\over \langle21\rangle}
\nonumber\\ && \biggl\{ - \langle 4|1|3\rangle\ 
(2s_{12}H_1 + \Delta H_{12}) 
- \langle 4|2|3\rangle\ (2s_{12}H_2 + \Delta H_{10}) \nonumber\\ && \quad
+2 ( \langle 4|1|3\rangle s_{24} + \langle 4|2|3\rangle s_{13} )\ H_4
\nonumber\\ && 
\quad + 2 \bigl[ \langle 4|1|3\rangle S_2 + \langle 4|2|3\rangle S_1
- (\langle 4|1|3\rangle s_{24} + \langle 4|2|3\rangle s_{13} ) \bigr]\
H_5 \nonumber\\ && 
+ {4\over s_{13} s_{2\h} } \biggl[ 
\left[ s_{13} (2s_{12} + S_2) - s_{24} s_{34} \right]
( \langle 4|1|3\rangle s_{24} + \langle 4|2|3\rangle s_{13} )
\nonumber\\ && \qquad + \left[ 2 s_{24} s_{34} (s_{12} + S_2 )
- s_{13} (s_{12} + S_2 )^2 \right]\ \langle 4|1|3\rangle
\nonumber\\ && \qquad + s_{13} (s_{34} S_2 - s_{12} S_1)
\langle 4|2|3\rangle \biggr] A_1(2;134) \nonumber\\ && 
+ {4\over s_{24} s_{1\h} } \biggl[
\left[ s_{24} (2s_{12} + S_1) - s_{13} s_{34} \right]
( \langle 4|1|3\rangle s_{24} + \langle 4|2|3\rangle s_{13} )
\nonumber\\ && \qquad + \bigl[ s_{24} \left[ s_{12} s_{34} +
(s_{12} + S_1) (s_{34} - S_2) \right] + s_{13} s_{34} S_2 \bigr]\
\langle 4|1|3\rangle \nonumber\\ && \qquad +
\left( s_{13} s_{34} S_1 + s_{24} s_{12}^2 \right) \langle 4|2|3\rangle
\biggr] A_1(1;234) \nonumber\\ && + 2 (S_1 - S_2)\
( \langle 4|1|3\rangle + \langle 4|2|3\rangle )\ A_1(12;34)
\nonumber\\ && - 4 
( \langle 4|1|3\rangle - \langle 4|2|3\rangle )\ A_2(12;34) \biggr\}\
+\ \ord(s^0)\ .\label{appiii}
\eea

\section{The large top-mass limit}
\label{sec:toplim}

In the large top-mass limit, $\mt^2\gg \mh^2/4$, the
gluon-Higgs coupling via a top-quark loop is given by a low-energy theorem 
through an effective Lagrangian~\cite{Shifman:1979eb,Ellis:1976ap},
\beq
\label{efflag}
\cL_\eff = {1\over 4}\, A H \, G^A_{\mu \nu} G^{A\,\mu \nu} \(1+\frac{11}{4}
\frac{\as}{\pi} + {\cal O}\(\as^2\)\)  \, , 
\eeq
where $G^A_{\mu \nu}$ is the field strength of the gluon field and $H$ is the
Higgs-boson field and the effective coupling $A$ is given in
Eq.~(\ref{eq:A_def}).

\subsection[Color decomposition of the amplitudes for Higgs
+ ${n}$ partons]{Color decomposition of the amplitudes for Higgs
+ $\boldsymbol{n}$ partons} 
\label{sec:hng}

Since the Higgs boson is a color singlet, the color structure of
the QCD amplitudes for Higgs $ {} + n$ partons in the large top-mass limit
is the same as the one of the tree $n$-parton amplitudes~\cite{Mangano:1990by}.
We repeat it here for later convenience.

The color decomposition of the tree $n$-gluon amplitudes 
is~\cite{Mangano:1990by,DelDuca:2000ha,DelDuca:2000rs}~\footnote{The 
normalisation $\tr(T^a T^b) = \delta^{ab}/2$
is the origin of the factor $2^{n/2}$ which we make explicit
rather than carrying it over in the sub-amplitudes,
as in Ref.~\cite{DelDuca:2000ha}. An additional factor 2 is pulled out
in order to use the same normalisation of the sub-amplitudes as in 
Ref.~\cite{Dawson:1992au}.} 
\begin{eqnarray}
\lefteqn{ \cM_n(1,\ldots,n) }\nonumber\\ &=& 2^{(n-2)/2} g^{n-2} 
\sum_{\sigma\in S_n/{\mathbb Z}_n}
   \tr \(T^{a_{\sigma_1}}\ldots T^{a_{\sigma_n}}\)\,
    m_n\(\sigma_1,\ldots,\sigma_n\) \label{GluonDecompNew}\\
&=& 2^{(n-2)/2} \frac{(ig)^{n-2}}{2} \sum_{\sigma\in S_{n-2}}
      f^{a_1 a_{\sigma_2} x_1} f^{x_1 a_{\sigma_3} x_2} \ldots f^{x_{n-3} 
a_{\sigma_{(n-1)}} a_n} 
m_n\(1,\sigma_2,\ldots,\sigma_{n-1},n\)\, ,\nn
\end{eqnarray}
where $S_n/{\mathbb Z}_n$ are the non-cyclic permutations of the $n$ gluons.
The dependence on the particle helicities and momenta in the sub-amplitude,
and on the gluon colors in the trace, is implicit
in labelling each leg with the index $i$.
Helicities and momenta are defined as if all particles were outgoing.
The gauge invariant sub-amplitudes $m$ are invariant under a cyclical
permutation of the arguments, and acquire a factor $(-1)^n$ under a
{\it reflection}, {\it i.e.} when the arguments are taken in reverse
order~\cite{Berends:1988me}.

The color decomposition of the tree amplitudes for $(n-2)$ gluons 
and a $q\bar q$ pair is
\begin{equation}
 \cM_n(q,\bar{q}; 3,\ldots,n)
  = 2^{(n-4)/2} g^{n-2} \sum_{\sigma\in S_{(n-2)}}
   (T^{a_{\sigma_3}}\ldots T^{a_{\sigma_n}})_{i}^{~\jb}\
    m_n(1_q,2_{\bar{q}}; \sigma_3,\ldots,\sigma_n)\, ,
\label{TwoQuarkGluonDecomp}
\end{equation}
where $S_{(n-2)}$ is the permutation group of the $(n-2)$ gluons.
Reflection symmetry is the same as for gluons only, for gluons and/or quarks
alike. 

\subsection{Sub-amplitudes for Higgs + three partons}
\label{sec:toplimhj}

The color decomposition of the tree amplitudes for Higgs plus three gluons 
is given in \eqn{GluonDecompNew}, with $n$ = 3. The independent
sub-amplitudes are~\cite{Kauffman:1997ix}
\begin{eqnarray}
&&m_3(1^+,2^+,3^+)= i A {\mh^4\over \la 1 2\ra \la 2 3\ra
   \la 3 1\ra} \, ,\label{ppp} \\  
&&m_3(1^-,2^+,3^+)= i A {[23]^3 \over [1 2] [13]}\, ,
\label{mpp}
\end{eqnarray}
with spinor products and currents defined in \app{sec:appa}.
All of the other sub-amplitudes can be obtained by relabelling and
by use of reflection symmetry, and parity inversion.
Parity inversion flips the helicities of all particles,
and it is accomplished by the substitution
$\spa{i}.j \leftrightarrow \spb{j}.i$.

The color decomposition of the tree amplitudes for Higgs, a gluon
and a $q\bar q$ pair is given in \eqn{TwoQuarkGluonDecomp} for $n$ = 3.
There is only one independent sub-amplitude~\cite{Kauffman:1997ix}
\beq
m_3(1_q^-, 2_{\bar{q}}^+; 3^+) = 
i A  { [23]^2 \over [12] }\, .\label{qmpp}
\eeq
All other sub-amplitudes can be obtained by use of parity inversion and charge 
conjugation~\footnote{In performing parity inversion, there is
a factor of $-1$ for each pair of quarks participating in the amplitude.}.
Following the conventions of Ref.~\cite{DelDuca:2000pa}, charge conjugation 
swaps quarks and antiquarks without inverting helicities.

\subsection{Sub-amplitudes for Higgs + four partons}
\label{sec:toplimhjj}

The color decomposition of the amplitudes for Higgs plus four gluons 
in the large top-mass limit
is given in \eqn{GluonDecompNew} for $n$ = 4. The independent
sub-amplitudes are~\cite{Dawson:1992au}
\begin{eqnarray}
&&m_4(1^+,2^+,3^+,4^+)= i A {\mh^4\over \la 1 2\ra \la 2 3\ra
   \la 3 4\ra \la 4 1\ra} \, ,\label{pppp}\\ 
&&m_4(1^-,2^+,3^+,4^+) = i A \left\{
 {\la 1|\ph |3\ra^2 [24]^2 \over s_{124} s_{12} s_{14}}     
+ {\la 1|\ph | 4 \ra^2 [2 3]^2 \over s_{123} s_{12} s_{23}} 
+ {\la 1|\ph | 2\ra^2 [3 4]^2 \over s_{134} s_{14} s_{34}} \right. \nonumber\\
&&\qquad  \left. - {[2 4] \over 
[ 1 2 ] \la 2 3\ra \la 3 4 \ra   [ 4 1]} 
\biggl[ s_{23} {\la 1 |\ph | 2 \ra \over \la 4 1\ra } + s_{34} 
{\la 1 |\ph | 4 \ra \over \la 1 2\ra }
-[2 4 ] s_{234}\biggr]  \right\} \, ,
\label{mppp}\\ 
&&m_4(1^-,2^-,3^+,4^+)= iA \left( {\la 1 2\ra^4\over \la 12 \ra 
\la 23 \ra \la 34\ra \la 41\ra}
 + {[34]^4 \over [1 2] [23] [34] [41]} \right) \, ,
\label{mmpp}
\end{eqnarray}
where $s_{ijk} = \(p_i+p_j+p_k\)^2$.
All of the other sub-amplitudes can be obtained by relabelling and
by use of reflection symmetry, and parity inversion.

The color decomposition of the amplitudes for Higgs, two gluons
and a $q\bar q$ pair is given in \eqn{TwoQuarkGluonDecomp}, with $n$ = 4.
The independent sub-amplitudes are~\cite{Kauffman:1997ix}~\footnote{For
the color ordering on the fermion line we choose the convention of 
Ref.~\cite{Mangano:1990by},
which is the opposite of the one used in Ref.~\cite{Kauffman:1997ix}.}
\bea
m_4(1^+, 2^+; 3_q^+, 4_{\bar{q}}^-) &=&  i A \left[
{ \la 4| \ph| 2 \ra^2 \over s_{341} }
  { [31] \over \la 41 \ra } \left({1\over s_{34}} + {1\over s_{31}} \right)
  - { \la 4| \ph| 1 \ra^2 \over s_{342} s_{34}} { [32] \over \la 42 \ra }
\right.
\nn\\
  && \left.\phantom{- i A \Bigg[}
- { \la 4| \ph| 3 \ra^2 \over [34] \la 41 \ra \la 42 \ra \la 12 \ra }
\right]\, ,  \label{qpmpp} \\
m_4(1^-, 2^+; 3_q^+, 4_{\bar{q}}^-) &=& -i A \left(
  {\la 41\ra^3 \over \la 34\ra \la 42\ra \la 12\ra } 
  - { [32]^3 \over [34][31][12] } \right) \, , \label{qpmmp} \\
m_4(1^+, 2^-; 3_q^+, 4_{\bar{q}}^-) &=& -i A \left( -
{ [31]^2 [41] \over [34][42][12] }
  + {\la 32\ra \la 42\ra^2 \over \la 34\ra \la 31\ra \la 12\ra } \right).
\label{qpmpm}
\eea
All of the other sub-amplitudes can be obtained by relabelling and
by use of parity inversion, reflection symmetry and charge 
conjugation.

The color decomposition of the amplitudes for Higgs and two different $q\bar
q$ pairs is given in \eqn{FourQuarkDecomp}.  There is one independent
sub-amplitude~\cite{Kauffman:1997ix}
\beq
m_4(1_q^+, 2_{\bar{q}}^-; 3_Q^+, 4_{\Qb}^-) = -i A \left(
{\la 24\ra^2 \over \la 12\ra \la 34\ra} + { [13]^2 \over [12][34] } \right) \, 
.\label{hqqqqt}
\eeq
All the other sub-amplitudes can be obtained by relabelling and
by use of parity inversion, reflection symmetry and charge 
conjugation.


\end{document}